\documentclass[aps,prl,twocolumn,superscriptaddress,showpacs]{revtex4-1}

\usepackage[utf8]{inputenc}
\usepackage[T1]{fontenc}

\usepackage{graphicx}  % needed for figures
\usepackage{dcolumn}   % needed for some tables
\usepackage{bm}        % for math
\usepackage{amssymb}   % for math
\usepackage{amsmath}
\usepackage{units}
\usepackage[english]{babel}
\usepackage[dvipsnames]{xcolor}
\usepackage{xfrac}
\usepackage{multirow}

\usepackage[%
colorlinks=true,
urlcolor=RoyalBlue,
linkcolor=RoyalBlue,
citecolor=RoyalBlue,
]{hyperref}

\usepackage{vmargin}
\setpapersize{A4} \oddsidemargin20mm \textwidth170mm \topmargin10mm
\usepackage{sidecap}
\sidecaptionvpos{figure}{t}  

\usepackage[type1]{libertine}                                        % Linux Libertine für zweispaltige Texte
\usepackage{textcomp}% Required to get special symbols
\usepackage[scaled=.85]{beramono}% Typewriter font
\usepackage[libertine,cmintegrals,cmbraces,vvarbb,slantedGreek]{newtxmath}
\usepackage[scr=boondoxo]{mathalfa}% Extra math symbols
\usepackage{bm}% Extra bold faces
\usepackage[lf]{carlito}

\usepackage{braket}
\usepackage{bm}

\makeatletter
\DeclareRobustCommand{\cev}[1]{%
  \mathpalette\do@cev{#1}%
}
\newcommand{\do@cev}[2]{%
  \fix@cev{#1}{+}%
  \reflectbox{$\m@th#1\vec{\reflectbox{$\fix@cev{#1}{-}\m@th#1#2\fix@cev{#1}{+}$}}$}%
  \fix@cev{#1}{-}%
}
\newcommand{\fix@cev}[2]{%
  \ifx#1\displaystyle
    \mkern#23mu
  \else
    \ifx#1\textstyle
      \mkern#23mu
    \else
      \ifx#1\scriptstyle
        \mkern#22mu
      \else
        \mkern#22mu
      \fi
    \fi
  \fi
}
\makeatother

\newcommand{\spinhalf}{spin-\sfrac{1}{2}}

\newcommand{\panel}[1]{(#1)}
\newcommand{\panelcaption}[1]{(#1)}
\newcommand{\panelsubcaption}[1]{(#1)}

\begin{document}

\title{Tracking a Spin-Polarized Superconducting Bound State across a Quantum Phase Transition}

\author{Sujoy Karan}
\email{s.karan@fkf.mpg.de}
\affiliation{Max Planck Institute for Solid State Research, Heisenbergstra{\ss}e 1, 70569 Stuttgart, Germany}

\author{Haonan Huang}
\affiliation{Max Planck Institute for Solid State Research, Heisenbergstra{\ss}e 1, 70569 Stuttgart, Germany}

\author{Alexander Ivanovic}
\affiliation{Institute for Complex Quantum Systems and IQST, Universität Ulm, Albert-Einstein-Allee 11, 89069 Ulm, Germany}

\author{Ciprian Padurariu}
\affiliation{Institute for Complex Quantum Systems and IQST, Universität Ulm, Albert-Einstein-Allee 11, 89069 Ulm, Germany}

\author{Bj\"orn Kubala}
\affiliation{Institute for Complex Quantum Systems and IQST, Universität Ulm, Albert-Einstein-Allee 11, 89069 Ulm, Germany}
\affiliation{Institute for Quantum Technologies, German Aerospace Center (DLR), Wilhelm-Runge-Straße 10, 89081 Ulm, Germany}

\author{Klaus Kern}
\affiliation{Max Planck Institute for Solid State Research, Heisenbergstra{\ss}e 1, 70569 Stuttgart, Germany}
\affiliation{Institut de Physique, Ecole Polytechnique F{\'e}d{\'e}rale de Lausanne, 1015 Lausanne, Switzerland}

\author{Joachim Ankerhold}
\affiliation{Institute for Complex Quantum Systems and IQST, Universität Ulm, Albert-Einstein-Allee 11, 89069 Ulm, Germany}

\author{Christian R. Ast}
\email{c.ast@fkf.mpg.de}
\affiliation{Max Planck Institute for Solid State Research, Heisenbergstra{\ss}e 1, 70569 Stuttgart, Germany}

\date{\today}

\begin{abstract}
The magnetic exchange coupling between magnetic impurities and a superconductor induce so-called Yu-Shiba-Rusinov (YSR) states which undergo a quantum phase transition (QPT) upon increasing the exchange interaction beyond a critical value. While the evolution through the QPT is readily observable, in particular if the YSR state features an electron-hole asymmetry, the concomitant change in the ground state is more difficult to identify. We use ultralow temperature scanning tunneling microscopy to demonstrate how the change in the YSR ground state across the QPT can be directly observed for a \spinhalf\ impurity in a magnetic field. We observe a change in the excitation spectrum from the doublet ground state in the free spin regime (two spectral features) to the singlet ground state in the screened spin regime (four spectral features). We also identify a transition regime, where the YSR excitation energy is smaller than the Zeeman energy. We thus provide a straightforward way for unambiguously identifying the ground state of a \spinhalf\ YSR state.
\end{abstract}

\maketitle

Unpaired spins in impurities coupled to a superconductor induce discrete sub-gap excitations, the Yu-Shiba-Rusinov (YSR) states \cite{yu_bound_1965,shiba_classical_1968,rusinov_superconductivity_1969,yazdani_probing_1997}, through an exchange interaction produced locally via impurity-superconductor coupling. If the exchange coupling increases beyond a critical value the YSR states undergo a quantum phase transition (QPT) such that the initially free spin becomes screened \cite{balatsky_impurity-induced_2006,heinrich_single_2018}. The transition through a QPT has been attributed to a reversal in the asymmetry of the spectral weight of electron and hole excitation components, which are readily observed in a scanning tunneling microscope (STM) \cite{farinacci_tuning_2018,kuster_long_2021,franke_competition_2011,kamlapure_correlation_2021,malavolti_tunable_2018,bauer_microscopic_2013,chatzopoulos_spatially_2021}. This reversal in spectral weight holds, however, only in the simplest approximation that all higher order effects are ignored. The spectral weight does not reflect the particle-hole asymmetry if, for example, the system is already in the resonant Andreev reflection regime \cite{ruby_tunneling_2015} or tunneling paths are interfering \cite{farinacci_interfering_2020}. Most crucially, it is \textit{a priori} not possible with the STM to identify to which side of the quantum phase transition the system belongs.

A straightforward albeit indirect and not entirely unambiguous way to manipulate the ground state of an atomic scale YSR resonance is to change the impurity-substrate coupling if the YSR impurity is susceptible to the atomic forces acting between tip and sample in the STM tunnel junction \cite{farinacci_tuning_2018,brand_electron_2018,malavolti_tunable_2018,kezilebieke_observation_2019,huang_quantum_2020}. The ambiguity arises because it is not \textit{a priori} clear whether the impurity-substrate coupling increases or decreases upon reducing the tip-sample distance. This calls for an unambiguous manifestation going beyond auxiliary measurements \cite{huang_quantum_2020} to distinguish the ground state of the YSR excitation.

\begin{SCfigure*}
\includegraphics[width=0.65\textwidth]{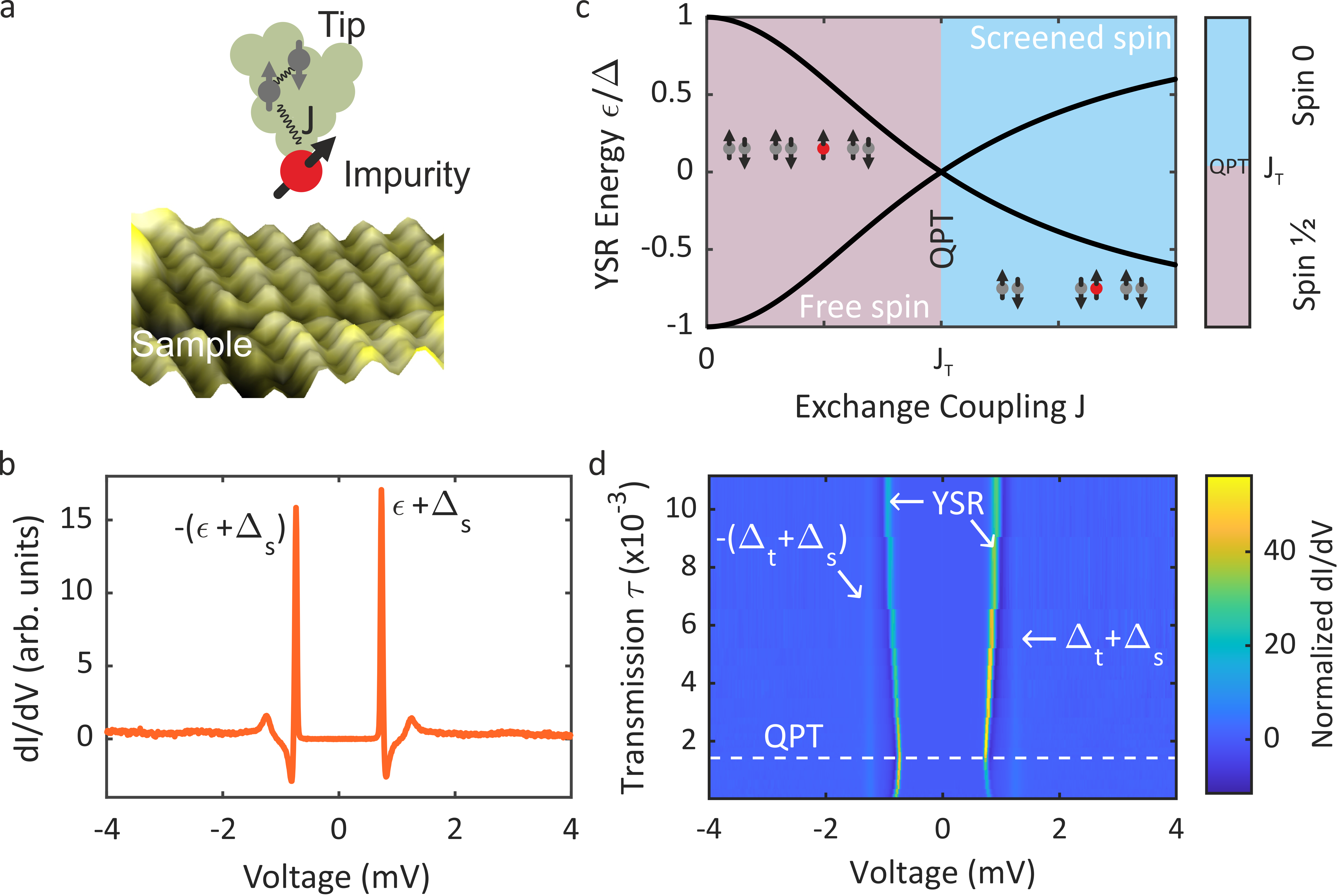}
\caption{\textbf{YSR state in the vicinity of a QPT} \panelcaption{a} Schematic of the tunnel junction incorporating a magnetic impurity at the tip apex. \panelcaption{b} Differential conductance spectrum at zero field showing the impurity-induced YSR states at $eV=\pm(\varepsilon+\Delta_\text{s})$. \panelcaption{c} The YSR excitation energy $\varepsilon$ vs.\ magnetic exchange coupling $J$. At the crossing of the YSR energies, the system undergoes a quantum phase transition (QPT) from a free spin doublet into a screened spin singlet state (see inset). \panelcaption{d} Normalized differential conductance spectra as function of junction transmission $\tau$. The QPT occurs, when the YSR peaks are closest to zero. The YSR peak crossing is not directly visible because both tip and sample are superconducting shifting the YSR peaks by the sample gap $\pm\Delta_s$. \panelcaption{b} and \panelcaption{d} The coherence peaks are visible at the sum of the tip and sample gap $eV=\pm(\Delta_t+\Delta_s)$. 
}
\label{fig:figure1}
\end{SCfigure*}

An independent observation identifying the ground state of the system across the QPT can be made by placing the YSR state in a Josephson junction (0-$\pi$ transition) \cite{karan_superconducting_2022}. Also, the zero-field splitting of YSR excitations due to effective anisotropic interactions in high-spin systems has been used to assign the ground state of different molecules on either side of the QPT  \cite{hatter_magnetic_2015}. While different YSR states have been studied with the STM in the presence of a magnetic field \cite{zitko_effects_2011,cornils_spin-resolved_2017,schneider_atomic-scale_2021,machida_zeeman_2022}, a continuous evolution of the YSR state across the QPT in a magnetic field has not been observed like in mesoscopic systems \cite{lee_spin-resolved_2014}. The challenge in observing a sizeable Zeeman splitting in a YSR state lies with the typically rather small critical magnetic field that quenches superconductivity. Here, we circumvent this problem by placing the YSR state at the tip apex \cite{karan_superconducting_2022,huang_tunnelling_2020}, where the superconductor is dimensionally confined, such that the critical field is considerably enhanced (Meservey-Tedrow-Fulde (MTF) effect) \cite{meservey_magnetic_1970,chen_vortices_2008,eltschka_probing_2014}. We use an ultralow temperature STM at 10\,mK to reduce the thermal energy much below the Zeeman energy and trace the spectral signatures associated with the changes in the YSR ground state across the QPT by continuously changing the impurity-substrate coupling (see Fig.\ \ref{fig:figure1}\panel{a}). 

A typical spectrum measured with a YSR functionalized superconducting vanadium tip on a superconducting V(100) sample at 10\,mK is shown in Fig.\ \ref{fig:figure1}\panel{b}. The electron and hole parts of the YSR state with energy $\varepsilon$ appear at a bias voltage $eV=\pm(\varepsilon+\Delta_\text{s})$ as prominent peaks. Due to the superconducting sample, the YSR peaks shift by the sample gap $eV=\pm\Delta_\text{s}$ away from zero bias voltage. The coherence peaks at $eV=\pm(\Delta_t+\Delta_s)$, which is the sum of the tip and sample gaps, are small indicating a dominant transport channel through the YSR state. We change the impurity-substrate coupling by varying the tip-sample distance, which modifies the atomic force acting on the impurity \cite{ternes_force_2008,ternes_interplay_2011}. This concomitantly changes the exchange coupling $J$ causing an evolution of the YSR energy $\varepsilon$ as shown in Fig.\ \ref{fig:figure1}\panel{c}. At a critical exchange coupling $J_\text{T}$, when the YSR energy is at zero, the system moves across a QPT such that the free impurity-spin ($J<J_\text{T}$) becomes screened ($J>J_\text{T}$) bringing about a change in the fermionic parity of the ground state \cite{sakurai_comments_1970}. This scenario is schematically depicted in the insets of Fig.\ \ref{fig:figure1}\panel{c}, where a doublet ($S=\sfrac{1}{2}$) transforms into a singlet ($S=0$) leading to the screening of the impurity spin.

Figure\ \ref{fig:figure1}\panel{d} shows how the YSR peaks evolve with the junction transmission $\tau=G_N/G_0$ ($G_N$: normal state conductance; $G_0 =2e^2/h$: conductance quantum with $e$ being the elementary charge and $h$ Planck's constant). The YSR peaks evolve continuously reaching the bias voltage closest to zero at the QPT. Because of the shift of the YSR state by the superconducting gap $\Delta_\text{s}$ of the other electrode (the substrate in this case) in the conductance spectrum, the zero crossing at the QPT is not observed directly. An inversion of the asymmetry in the YSR peak intensities is clearly visible, when the electron and hole excitation components switch sides across the QPT. However, it is not possible to judge from the tunneling spectra alone, on which side of the QPT the system is. 

Turning on a magnetic field, the $S=\sfrac{1}{2}$-state splits into two levels. In the free spin regime, the spin down state (see Fig.\ \ref{fig:figure3}\panel{a}) turns into the non-degenerate ground state.  Its higher lying spin-flipped partner is thermally not populated due to the extremely low temperature of 10\,mK. Only the screened $S=0$-state appears as a transport channel lying energectically above the doublet. Since it does not change in the magentic field, it induces only one spectral feature on either side of the Fermi level. In contrast, beyond the QPT, the $S=0$-state becomes the ground state and charge transfer is possible through the spin-doublet (details see below). This can be seen in Fig.\ \ref{fig:figure2}\panel{a}, which shows two representative differential conductance spectra on either side of the QPT at a magnetic field of 750\,mT. The orange spectrum shows two features (one on either side of the Fermi level), which indicates that the system is in the free spin regime. The sample is already normal conducting at 750\,mT, such that there is no shift of the YSR peak by $\Delta_\text{s}$. The YSR tip is still superconducting due to the MTF effect. In the screened spin regime, ground state and excited state are interchanged, such that now two transitions into the upper and lower Zeeman split $S=\sfrac{1}{2}$ levels are possible from the single ground state $S=0$ level. As a result, the spectrum measured in the screened spin regime (the blue curve) shows four spectral features (two on either side of the Fermi level). This distinction is only possible, if the Zeeman energy is much larger than the thermal energy. If this is not the case, two spectral features will be visible on either side of the QPT and a more detailed analysis of the spectral weight has to be done to distinguish the ground states \cite{hatter_magnetic_2015}. 

As has been demonstrated before \cite{huang_quantum_2020,huang_tunnelling_2020,karan_superconducting_2022,farinacci_tuning_2018,kuster_long_2021,franke_competition_2011,kamlapure_correlation_2021,malavolti_tunable_2018,bauer_microscopic_2013,chatzopoulos_spatially_2021}, we exploit the changing atomic forces in the tunnel junction when reducing the tip-sample distance to change the impurity-superconductor coupling thereby shifting the YSR state energy. We note that depending on the particular system, the impurity-substrate coupling can increase or decrease during tip approach. The evolution of the YSR state through the QPT for two different magnetic fields are shown in Figs.\ \ref{fig:figure2}\panel{b} and \panel{c} as function of the tunnel junction transmission $\tau$ (i.e.\ junction conductance). Here, we can see directly that the screened spin regime featuring four spectral peaks is at higher transmissions and the free spin regime featuring only two spectral peak is at lower transmissions. This actually implies that the impurity-superconductor coupling increases with increasing transmission, which is verified by an additional analysis of the Kondo effect at higher magnetic fields below. The data in Fig.\ \ref{fig:figure2}\panel{b} was taken at 750\,mT, which results in a stronger Zeeman splitting than the data in Fig.\ \ref{fig:figure2}\panel{c}, which was taken at 500\,mT having less Zeeman splitting. Still, both data sets show qualitatively the same behavior across the QPT as expected from the discussion above. 

\begin{SCfigure*}
\centering
\includegraphics[width=0.68\textwidth]{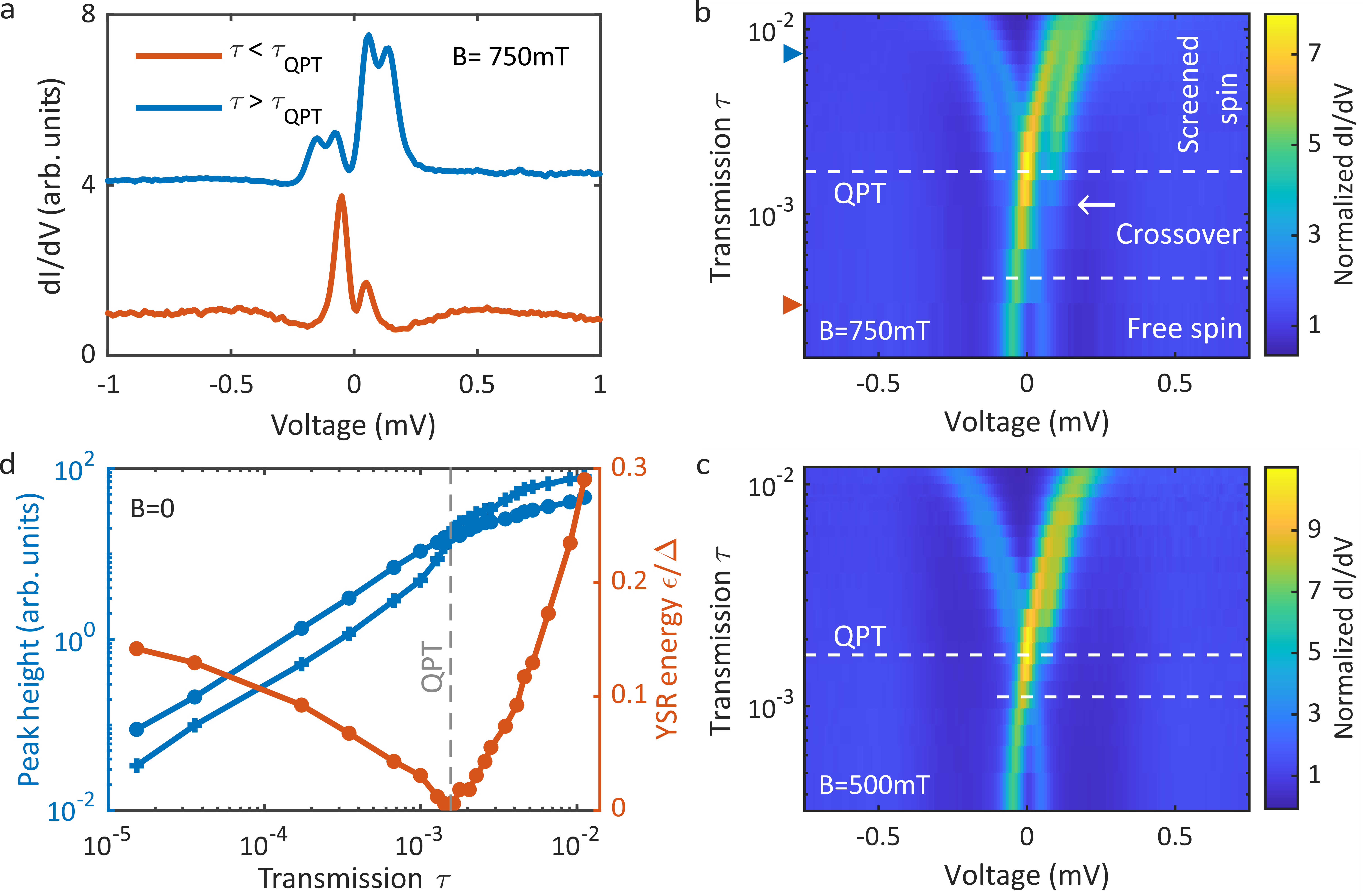}
\caption{\textbf{Magnetic field dependence of YSR states across QPT.}
\panelcaption{a} Differential conductance spectra at a magnetic field of $B=750$\,mT at two junction transmissions, one below and one above the QPT marked by arrows in panel \panelsubcaption{b}. \panelcaption{b} Differential conductance map at 750\,mT as function of junction transmission revealing the dispersion of the YSR states across the QPT. The sample is normal conducting, so that the spectral features cross at the Fermi level. \panelcaption{c} Same as \panelsubcaption{b} for $500$\,mT with a correspondingly reduced Zeeman splitting. \panelcaption{d} YSR peak heights (blue) at zero field of the left and right peak. The height inverts across the QPT, where the YSR energies $\varepsilon$ (orange) are zero.
}
\label{fig:figure2}
\end{SCfigure*}

\begin{figure}
\includegraphics[width=0.8\linewidth]{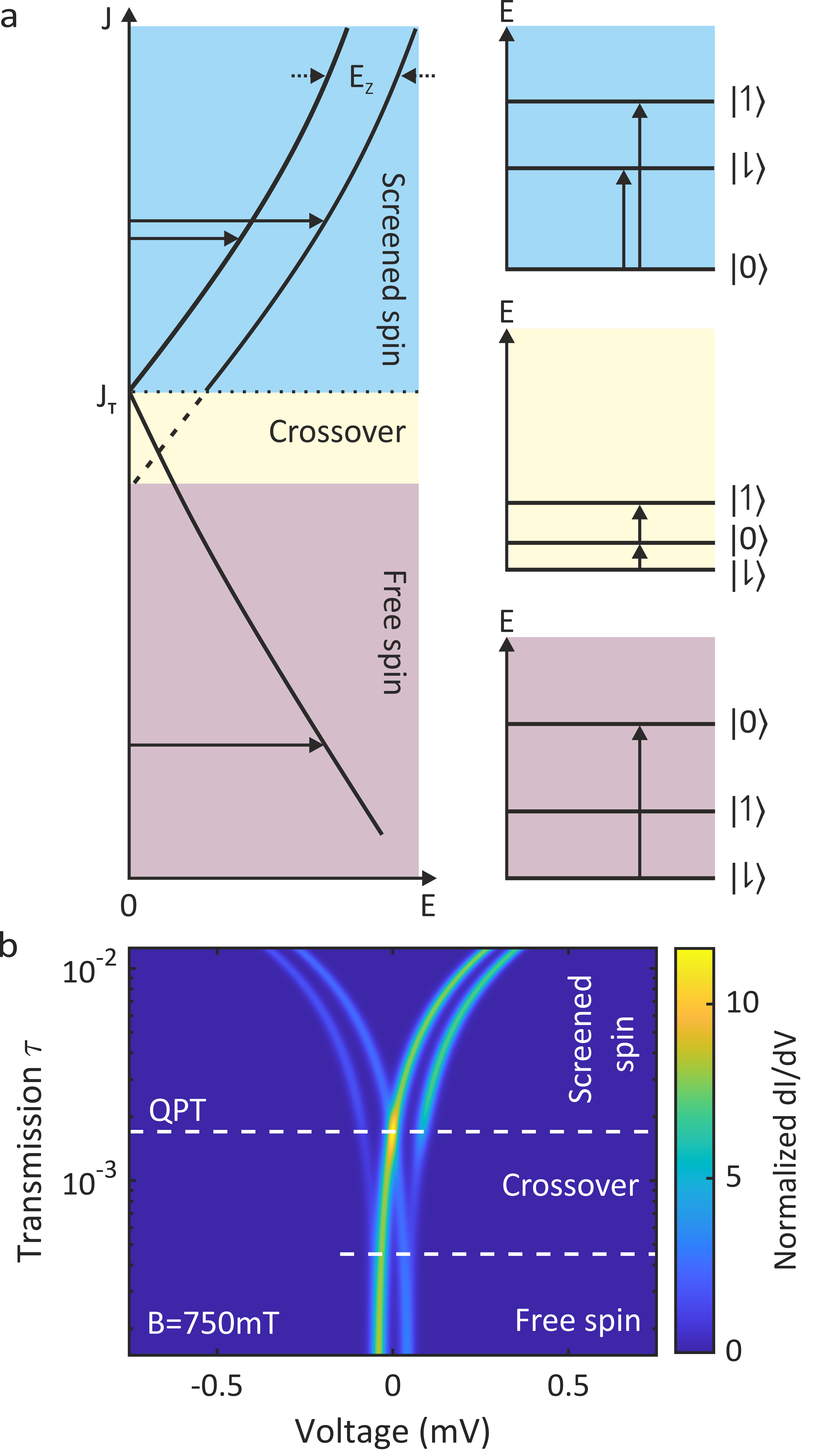}
\caption{\textbf{Theoretical modeling of the Zeeman splitting across the QPT.} \panelcaption{a} The sketches illustrate the level structures of a YSR state at finite magnetic field in different regimes across the QPT. The Zeeman effect lifts the spin degeneracy of the doublet state leading to two possible transitions in the screened spin regime (shaded blue). The YSR states are split by the Zeeman energy $E_z=g\mu_{B}B$, where $g$ is the $g$-factor and $\mu_\text{B}$, is the Bohr magneton. The crossover regime (shaded yellow) in the proximity of the QPT still allows for a second transition involving thermal excitation of the YSR state and two-electron tunneling processes. The free spin regime (shaded red) allows only for one transition. \panelcaption{b} Calculation of the differential conductance spectra across the QPT based on a master equation including single electron and two electron processes. The experimental data in Fig.\ \ref{fig:figure2}\panelsubcaption{b} is well reproduced.}
\label{fig:figure3}
\end{figure}

In addition to these two regimes, we found a crossover regime, where the two outer spectral features extend into the free spin regime, which is seen for both magnetic field values in Fig.\ \ref{fig:figure2}\panel{b} and \panel{c}. Due to the higher magnetic field in Fig.\ \ref{fig:figure2}\panel{b} than in \panel{c}, the crossover regime is also wider. The crossover regime marks a small region, where the excitation energy of the YSR state is smaller than the Zeeman splitting ($\varepsilon<E_\text{Z}$). The outer spectral feature (marked by the arrow in Fig.\ \ref{fig:figure2}\panel{b}) in the crossover regime is a combination of quasiparticle tunneling from the thermally excited YSR state, which becomes exponentially suppressed as the YSR energy increases, and two-electron tunneling processes, i.e.\ resonant Andreev processes (see below and Supplementary Information \cite{sinf}).

The different regimes for a \spinhalf\ impurity are schematically displayed in Fig.\ \ref{fig:figure3}\panel{a}. The screened spin regime (blue shade), where the exchange coupling is strong $J>J_\text{T}$, features an $S=0$ ground state and a Zeeman split excited $S=\sfrac{1}{2}$ state. Two transitions are possible ($\ket{0}\rightarrow\ket{\downharpoonright}$ and $\ket{0}\rightarrow\ket{\upharpoonleft}$) as shown on the right blue panel. Lowering the exchange coupling, the $\ket{\downharpoonright}$ state becomes the ground state at the QPT ($J=J_\text{T}$). Interestingly, in the crossover regime the excited state $\ket{0}$ is energetically between the ground state $\ket{\downharpoonright}$ and the Zeeman split state $\ket{\upharpoonleft}$. Therefore, both thermally excited tunneling and two-electron tunneling processes are possible resulting in the outer spectral feature (white arrow in Fig.\ \ref{fig:figure2}\panel{b}). As a consequence, two transitions can be observed ($\ket{\downharpoonright}\rightarrow\ket{0}$ and $\ket{0}\rightarrow\ket{\upharpoonleft}$). Further reducing the exchange coupling into the free spin regime reduces the visible transitions to one ($\ket{\downharpoonright}\rightarrow\ket{0}$), because the Zeeman split state $\ket{\upharpoonleft}$ cannot be thermally excited at 10\,mK. We can reproduce the experimental findings theoretically by calculating a tunneling current from a master equation involving both single electron and two electron processes (for details see the Supplementary Information \cite{sinf}). The calculation in Fig.\ \ref{fig:figure3}\panel{b} has been done for a magnetic field of 750\,mT comparable to the experimental data in Fig.\ \ref{fig:figure2}\panel{b}. All the features that we observed experimentally are reproduced in the calculations.

\begin{figure}
\includegraphics[width=0.99\linewidth]{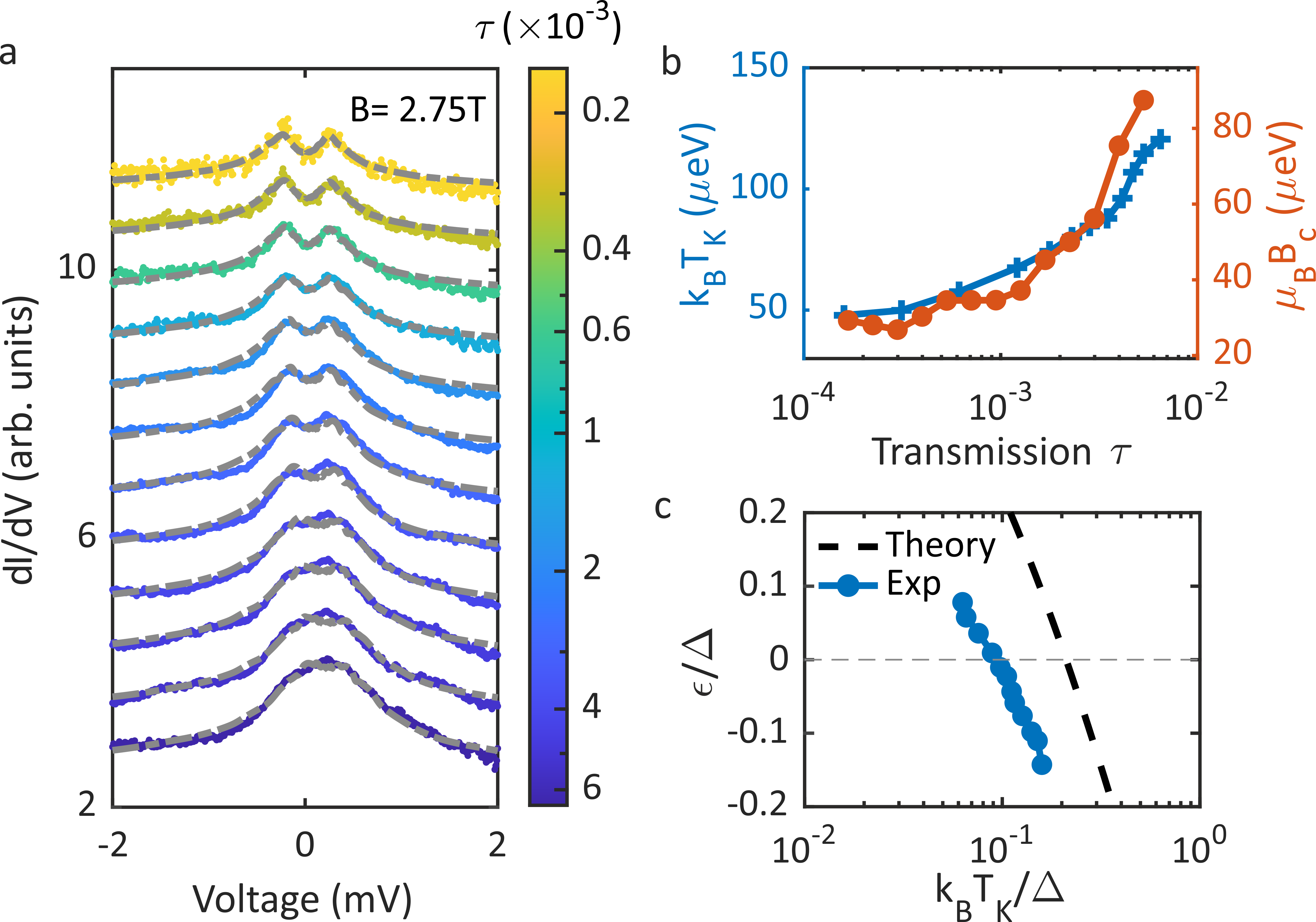} 
\caption{\textbf{Kondo effect and numerical renormalization group (NRG) analysis}
\panelcaption{a} Differential conductance spectra featuring a Kondo peak at different junction transmissions $\tau$. Dashed gray lines represent fits of the data to NRG calculations \cite{zitko_energy_2009}. \panelcaption{b} The Kondo temperature extracted from the fit in \panelsubcaption{a} as function of transmission $\tau$ in blue along with the critical field $B_\text{c}$, above which the Kondo peak starts splitting. \panelcaption{c} YSR state energy vs.\ Kondo temperature both scaled to the superconducting gap $\Delta$. The dashed black curve represents the universal scaling predicted from the NRG model. 
}
\label{fig:figure4}
\end{figure}

In order to independently verify the evolution of the YSR state through the QPT, we take a closer look at the YSR peak height and the resulting Kondo effect in the normal conducting state. The evolution of the YSR peak height is plotted in Fig.\ \ref{fig:figure2}\panel{d} in blue for the left and right peak as function of junction transmission. In the same graph the YSR peak energy is shown in orange. At the QPT (vertical dashed line), the YSR energy is zero and the peak height reverses indicating the QPT. This reversal is observable so clearly because resonant Andreev processes have not yet become significant. Further, we increase the magnetic field to 2.75\,T such that both tip and sample become normal conducting and a Kondo peak appears \cite{kondo_resistance_1964,li_kondo_1998,madhavan_tunneling_1998,ternes_spectroscopic_2008}. This is shown in Fig.\ \ref{fig:figure4}\panel{a}, where the Kondo peak around zero bias voltage is displayed as a function of the junction transmission $\tau$. We already see that the splitting of the Kondo peak in the magnetic field decreases as the transmission increases, which indicates that the Kondo temperature increases with increasing transmission. The higher Kondo temperature implies a stronger screening, which means that the Kondo peak starts splitting at a higher critical magnetic field $B_\text{c}$. We have fitted the Kondo spectra using numerical renormalization group (NRG) theory \cite{huang_universal_2022}. This allows us to directly determine the Kondo temperature $T_\text{K}$ from the microscopic parameters extracted from the fit. The extracted Kondo temperature is shown in Fig.\ \ref{fig:figure4}\panel{b} as the blue line. It monotonously increases with increasing junction transmission $\tau$, which corroborates the previous finding that the exchange coupling increases with increasing transmission (cf.\ Fig.\ \ref{fig:figure2}\panel{b} and \panel{c}) \cite{huang_quantum_2020,cuevas_evolution_1998}. We also extracted the critical field $B_\text{c}$, where the Kondo peaks starts splitting, as a function of transmission. The values for the critical field $B_\text{c}$ are plotted in Fig.\ \ref{fig:figure4}\panel{b} as a red line. The critical field increases with increasing transmission and follows the Kondo temperature very well. This corroborates very well the increase in impurity-substrate coupling for an increasing junction transmission. We further find a relation of $k_\text{B}T_\text{K}=\alpha \mu_\text{B}B_\text{c}$ between the Kondo temperature and the critical field with $\alpha = 1.6$, which compares well with what has been found in the literature \cite{hewson_non-equilibrium_2005,zitko2009splitting,kretinin_spin-frac12_2011}. 

Furthermore, scaling the Kondo temperature $T_\text{K}$ and the YSR energy $\varepsilon$ to the superconducting gap $\Delta$, we compare the evolution across the QPT to the universal behavior predicted by NRG theory \cite{satori_numerical_1992,yoshioka_numerical_2000,bulla_numerical_2008}. The blue data points in Fig.\ \ref{fig:figure4}\panel{c} show the evolution of the YSR state across the QPT as function of the scaled Kondo temperature, which follows the predicted universal scaling (dashed line) with a slight offset. This deviation of the data from the universal curve is presumably due to subtle changes in the impurity-substrate coupling as a result of modifications in the atomic forces acting in the junction with and without the applied magnetic field. We, therefore, find a consistent picture for the behavior of the YSR state in a magnetic field across the quantum phase transition.

The evolution of the YSR state splitting across the QPT clearly demonstrates the change in the YSR ground state. For a \spinhalf\ system, the nature of the ground state can be straightforwardly identified simply by the number of peaks in the spectrum. For higher order spins, the situation remains simple as long as the system can be assumed to be magnetically isotropic \cite{machida_zeeman_2022}. If the system experiences a magnetic anisotropy, the analysis of the YSR states becomes more cumbersome \cite{zitko_effects_2011,oppen_yu-shiba-rusinov_2021}. Still, the evolution in a magnetic field as well as with changing impurity-superconductor coupling (if susceptible to the atomic forces of the tip) greatly facilitates the identification of the ground state maybe even the spin state itself. 

In summary, we present the evolution of a \spinhalf\ impurity derived YSR state in a magnetic field across the QPT. Due to the extremely low temperature of the STM, the change from a single feature spectrum (free spin regime) to a double feature spectrum (screened spin regime) is clearly visible. This allows for an unambiguous determination of the ground state of the YSR state. 

\section{Acknowledgments}

The authors thank Carlos Cuevas and Andrea Hofmann for fruitful discussions. This study was funded in part by the ERC Consolidator Grant AbsoluteSpin (Grant No. 681164). JA and CP gratefully acknowledge financial support by the IQST and the BMBF through QSens (project QComp).

\section{Methods}
The V(100) single crystal was sputtered (with Ar$^+$), annealed to about 925\,K, and cooled to ambient temperature repeatedly in ultra-high vacuum, ensuring an atomically flat sample surface.
Typical surface reconstructions form with oxygen diffused from the bulk \cite{koller_structure_2001,kralj_hraes_2003,huang_tunnelling_2020}. A small fraction of these defects exhibit YSR states \cite{huang_tunnelling_2020}. Similarly, we produce YSR states at the vanadium tip apex by repeatedly dipping the tip \textit{in situ} into the substrate\cite{karan_superconducting_2022,huang_tunnelling_2020}, which is verified in the conductance spectrum. This gives us full control to reproducibly design and define the junctions under investigation. We choose to use YSR functionalized tips for our experiments as they offered the flexibility to single out those fulfilling the required response to tip approach. Moreover, the YSR tips feature a range of YSR state energies and show a better junction stability at higher conductance than YSR states in the sample.

The experiments were performed in a low-temperature scanning tunneling microscope operating at 10\,mK. Differential tunneling conductance ($dI/dV$) spectra were recorded using an open feedback loop with a standard lock-in technique ($10\,\upmu\text{V}_\text{rms}$, 727.8\,Hz). In Fig.\ \ref{fig:figure4}\panel{a} a modulation amplitude of $50\,\upmu\text{V}_\text{rms}$ was used. The tunneling current was measured through the tip with the voltage bias applied to the sample. 

The calculations for the current based on the master equation are detailed in the Supplementary Information \cite{sinf}. For the Kondo spectra, we used the numerical renormalization group (NRG) theory in the framework of the single impurity Anderson model (SIAM) as implemented in ``NRG Ljubljana'' code \cite{zitko_energy_2009} to model the Kondo effect in a magnetic field. We fixed the Hubbard term $U=10$ to be much larger than the half bandwidth $D=1$ and modeled the asymmetry of the Kondo spectra using the intrinsic asymmetry parameter $\delta=\epsilon+U/2$ where $\epsilon$ is the impurity level \cite{huang_universal_2022}.  The best agreement with the experiment corresponds to $\delta=-2$. The only free parameter left for fitting the Kondo spectra is the impurity-substrate coupling $\Gamma$. The Kondo temperature was extracted from the fit through its definition with respect to the SIAM parameters \cite{schrieffer_relation_1966,yoshioka_numerical_2000,kadlecova2019practical} $k_\text{B}T_\text{K} = D_\text{eff}\sqrt{\rho J}\exp\left(-\frac{1}{\rho J}\right)\text{ with } \rho J=\frac{8\Gamma}{\pi U}\frac{1}{1-4(\delta/U)^2}$, in which the effective bandwidth satisfies $D_\text{eff}=0.182U\sqrt{1-4(\delta/U)^2}$ for $U\ll 1$ and $D_\text{eff}$ is a constant for $U\gg 1$ \cite{yoshioka_numerical_2000,zitko2007many,huang_universal_2022}.

\clearpage
\newpage

\onecolumngrid
\begin{center}
\textbf{\large Supplementary Material}
\end{center}
\vspace{1cm}
\twocolumngrid

\setcounter{figure}{0}
\setcounter{table}{0}
\setcounter{equation}{0}
\renewcommand{\thefigure}{S\arabic{figure}}
\renewcommand{\thetable}{S\Roman{table}}
\renewcommand{\theequation}{S\arabic{equation}}

\section{\label{sec:intro} Theory: Introduction}

A single \spinhalf\ impurity gives rise to an in-gap Yu-Shiba-Rusinov state (YSR). Based on the occupation of the YSR state, the system wavefunction can be a spin doublet, when the YSR state is unoccupied, or a singlet, in the opposite case \cite{si_Anderson1959,si_Zitko2011,si_Huang2020,si_Machida2022}.

The presence of a magnetic field splits the doublet states. We denote by $\ket{0\uparrow}$ and $\ket{0\downarrow}$ the spin doublet states, that correspond to the free impurity spin. The states indicate that the \spinhalf\ impurity is aligned, respectively anti-aligned, with the external magnetic field. The spin singlet, that corresponds to the screened impurity spin, is denoted by $\ket{1}$.
The notation adopted here differs from the notation in the main text, by adding an emphasis on the occupation of the YSR state, that we believe is more intuitive for transport calculations. The notation in the main text for the doublet states $\ket{\upharpoonleft}$ and $\ket{\downharpoonright}$, indicating the total spin state, is equivalent to the notation here $\ket{0\uparrow}$ and $\ket{0\downarrow}$, indicating an empty YSR state (free impurity spin) and the total spin. The notation in the main text for the singlet state $\ket{0}$, indicating the total spin $S=0$, is equivalent here to the state $\ket{1}$, indicating the occupation of the YSR state.

\begin{figure}
    \includegraphics[width=\columnwidth]{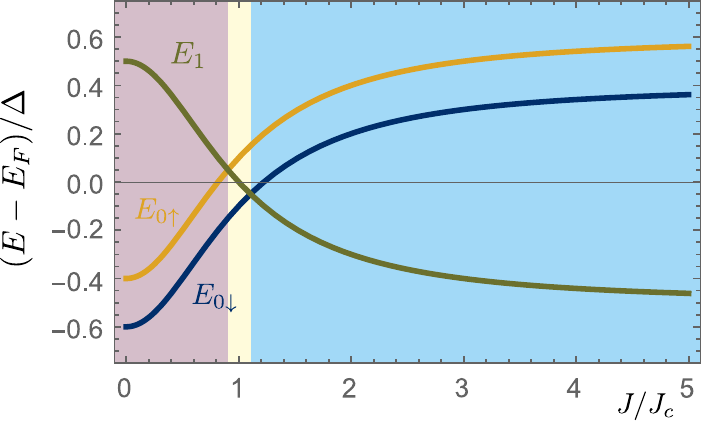}\hspace{1cm}\includegraphics[width=\columnwidth]{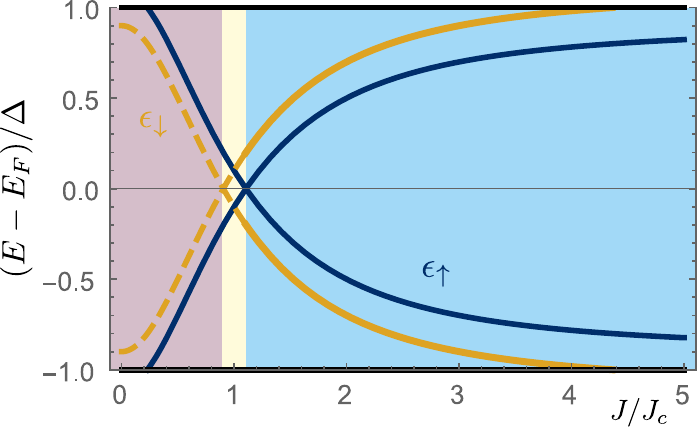}
    \caption{\label{fig:energy_diagrams} Energy of YSR states. Above: energies as a function of exchange coupling. The critical exchange coupling $J_c$ corresponds to the phase transition at zero magnetic field. The Zeeman energy is chosen as $E_Z=0.1\Delta$. Below: excitation energies as a function of exchange coupling. Both $\epsilon_\sigma$ and $-\epsilon_\sigma$ are shown, for $\sigma=\uparrow,\downarrow$. The dashed lines indicate the excitation energy involves two excited states, while the solid lines indicate excitations from the ground state. The background color indicates the free spin, crossover and screened spin regions, as in Fig. 3 of the main text.}
\end{figure}

The energies of the three states $E_{0\uparrow}$, $E_{0\downarrow}$, and $E_1$ depend on the exchange coupling strength and give rise to three regimes, as shown in Fig.~3 of the main text. The free spin regime $E_1>E_{0\uparrow},E_{0\downarrow}$, a crossover regime $E_{0\uparrow}>E_1>E_{0\downarrow}$, and the screened spin regime $E_{0\uparrow},E_{0\downarrow}>E_1$. We have assumed that $E_{0\uparrow}>E_{0\downarrow}$, with the energy difference determined by the Zeeman splitting $E_Z$ of the doublet state $E_{0\uparrow}-E_{0\downarrow}=E_Z$ (as illustrated in Fig.~\ref{fig:energy_diagrams}).

For transport calculations, it is useful to work with fermionic excitation energies. We define the energy required to add a quasiparticle with spin $\uparrow$ to the YSR state as $\epsilon_\uparrow = E_1-E_{0\downarrow}$. We note that the state involved is $\ket{0\downarrow}$, such that the impurity spin $\downarrow$ is screened by the added quasiparticle $\uparrow$. Similarly, we denote by $\epsilon_\downarrow = E_1-E_{0\uparrow}$, the energy required to add a quasiparticle with spin $\uparrow$ to the YSR state. Consistent with our assumptions above, we have $\epsilon_\uparrow-\epsilon_\downarrow=E_Z$, therefore $\epsilon_\uparrow>\epsilon_\downarrow$ (as shown in Fig.~\ref{fig:energy_diagrams}).

We note that it is possible to remove either a spin $\uparrow$, or a spin $\downarrow$ quasiparticle from the YSR state, when it is in the singlet state $\ket{1}$. The implication therefore is that transport processes can result in the flip of the impurity spin, e.g. $\ket{0\downarrow}\xrightarrow{\rm{add\:} \uparrow}\ket{1}\xrightarrow{\rm{remove\:} \downarrow}\ket{0\uparrow}$. Crucially, also the transport between the impurity and its host superconducting substrate can lead to such spin flip, as pointed out in Ref.~\cite{si_van_Gerven_Oei_2017}.

\subsection{Rate equations}

The simplest theoretical framework that captures the transport properties of the Zeeman split states takes the form of a rate equation for the probabilities to be in one of three states, $P_{0\uparrow}$, $P_{0\downarrow}$, and $P_{1}$. We denote by $\Gamma_{1\uparrow}$ ($\Gamma_{1\downarrow}$) the rate to \textit{remove} a quasiparticle with spin $\uparrow$ ($\downarrow$) from the YSR state, and by $\Gamma_{2\uparrow}$ ($\Gamma_{2\downarrow}$) the rate to \textit{add} a quasiparticle with spin $\uparrow$ ($\downarrow$) to the YSR state.

The rates represent a sum of all the contributing processes, intrinsic processes and tunneling processes, and depend on the bias voltage and the YSR state energy, that we parameterize by the exchange coupling strength $J$ (see Fig.~\ref{fig:energy_diagrams}).

The probabilities characterizing the three states obey the following rate equations,
\begin{align}
    \dot{P}_{0\downarrow}=&\Gamma_{1\uparrow}P_1-\Gamma_{2\uparrow}P_{0\downarrow},\\
    \dot{P}_{0\uparrow}=&\Gamma_{1\downarrow}P_1-\Gamma_{2\downarrow}P_{0\uparrow},\\
    \dot{P}_{1}=&\left(\Gamma_{2\uparrow}P_{0\downarrow}+\Gamma_{2\downarrow}P_{0\uparrow}\right)-(\Gamma_{1\uparrow}+\Gamma_{1\downarrow})P_1,
\end{align}
with the normalization condition $P_{0\uparrow}+P_{0\downarrow}+P_1=1$.

The steady state probabilities are given by
\begin{align}
    P_{0\uparrow} = \frac{\Gamma_{1\downarrow}\Gamma_{2\uparrow}}{D},\quad P_{0\downarrow}=\frac{\Gamma_{1\uparrow}\Gamma_{2\downarrow}}{D}, \quad P_1=\frac{\Gamma_{2\uparrow}\Gamma_{2\downarrow}}{D}.
    \label{eq:probabilities}
\end{align}
Where we have used the notation
\[D=\Gamma_{2\uparrow}\Gamma_{2\downarrow}+\Gamma_{1\uparrow}\Gamma_{2\downarrow}+\Gamma_{2\uparrow}\Gamma_{1\downarrow}.\]
In the following, we discuss the rates, which consist of intrinsic rates ($\Gamma^{(i)}_{1\sigma}$, $\Gamma^{(i)}_{2\sigma}$) and tunneling rates ($\Gamma^{(t)}_{1\sigma}$, $\Gamma^{(t)}_{2\sigma}$), with $\sigma\in\{\uparrow,\downarrow\}$, such that \[ \Gamma_{1\sigma}=\Gamma^{(i)}_{1\sigma}+\Gamma^{(t)}_{1\sigma}\quad \textrm{and} \quad \Gamma_{2\sigma}=\Gamma^{(i)}_{2\sigma}+\Gamma^{(t)}_{2\sigma}.\]

\subsection{Intrinsic rates}

The occupation of the YSR state can change due to intrinsic processes involving the quasiparticle population above the superconducting gap. We stress that these processes do not involve tunneling between the tip and substrate. The intrinsic process that describes the addition of a quasiparticle to the YSR state, changing the state from $\ket{0\sigma}$, with $\sigma\in\{\uparrow,\downarrow\}$, to the singlet state $\ket{1}$, requires that the quasiparticle has spin $\bar{\sigma}$, opposite to $\sigma$. We denote the corresponding rate by $\Gamma^{(i)}_{2\bar{\sigma}}$. Similarly, the rate to emit a quasiparticle with spin $\bar{\sigma}$ into the continuum, is denoted by $\Gamma^{(i)}_{1\bar{\sigma}}$. The latter process transforms the initial state $\ket{1}$ into the doublet state $\ket{0\sigma}$.

In absence of tunneling between tip and substrate, the intrinsic processes in the tip and are responsible for the equilibrium value of the YSR state occupation. We apply detailed balance to determine the relations between the intrinsic rates. The equilibrium population in each state in absence of tunneling $P^{(\rm{eq})}_{0\uparrow}$, $P^{(\rm{eq})}_{0\downarrow}$, and $P^{(\rm{eq})}_{1}$, are obtained from Eqs.~\ref{eq:probabilities} by setting the tunneling rates to zero, such that $\Gamma_{1\sigma}=\Gamma^{(i)}_{1\sigma}$ and $\Gamma_{2\sigma}=\Gamma^{(i)}_{2\sigma}$. We require that the equilibrium populations are related by the Fermi-Dirac distribution
\[
\frac{P^{(\rm{eq})}_1}{P^{(\rm{eq})}_{0\downarrow}}=\frac{\Gamma^{(i)}_{2\uparrow}}{\Gamma^{(i)}_{1\uparrow}}=\frac{n_F(\epsilon_\uparrow)}{1-n_F(\epsilon_\uparrow)},\quad \frac{P^{(\rm{eq})}_1}{P^{(\rm{eq})}_{0\uparrow}}=\frac{\Gamma^{(i)}_{2\downarrow}}{\Gamma^{(i)}_{1\downarrow}}=\frac{n_F(\epsilon_\downarrow)}{1-n_F(\epsilon_\downarrow)}.
\]
Furthermore, we will assume that the rate to emit a quasiparticle with spin $\sigma$ into the continuum, transition from $\ket{0\bar{\sigma}}$ to state $\ket{1}$, is independent of the orientation of the impurity spin $\bar{\sigma}$. Therefore, we have \[\Gamma^{(i)}_{1\uparrow}=\Gamma^{(i)}_{1\downarrow}=\Gamma^{(i)}_1.\]
This assumption is not necessary, but convenient to reduce the number of parameters of the model. The physical mechanism behind such intrinsic processes, as well as the origin of the quasiparticle population above the gap at mK temperatures, remain unknown.

\begin{figure}
    \includegraphics[width=\columnwidth]{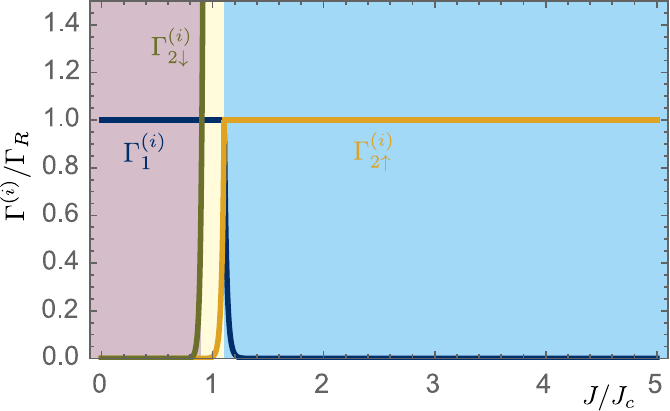}
    \caption{\label{fig:intrinsic} Intrinsic rates as a function of exchange coupling. The critical exchange coupling $J_c$ corresponds to the phase transition at zero magnetic field.}
\end{figure}

The two relations obtained by applying detailed balance, together with our assumption, express the intrinsic rates in terms of a single free parameter, which we denote $\Gamma_R$, an intrinsic rate of relaxation. We parameterize the intrinsic rates in terms of $\Gamma_R$, as follows
\begin{align}
    \Gamma^{(i)}_{1} =& \begin{cases}\Gamma_R, &\quad \epsilon_\uparrow \geq 0,\\
    \Gamma_R\exp(\epsilon_\uparrow/k_BT).
    &\quad \epsilon_\uparrow < 0\end{cases}\\
    \Gamma^{(i)}_{2\uparrow}=& \begin{cases}\Gamma_R \exp(-\epsilon_\uparrow/k_BT),
    &\quad \epsilon_\uparrow \geq 0,\\
    \Gamma_R,&\quad \epsilon_\uparrow < 0.\end{cases}\\
    \Gamma^{(i)}_{2\downarrow}=& \begin{cases}\Gamma_R \exp(-\epsilon_\downarrow/k_BT)
    ,&\quad \epsilon_\uparrow \geq 0,\\
    \Gamma_R\exp\left[(\epsilon_\uparrow-\epsilon_\downarrow)/k_BT)\right]
    ,&\quad \epsilon_\uparrow < 0.
    \end{cases}
\end{align}
The intrinsic rates are shown in Fig.~\ref{fig:intrinsic}. The parametrization chose ensures that the intrinsic rates $\Gamma^{(i)}_{1}$ and $\Gamma^{(i)}_{2\uparrow}$ are bound by $\Gamma_R$. The intrinsic rate $\Gamma^{(i)}_{2\downarrow}$ becomes much larger than $\Gamma_R$ in the regime $\epsilon_\uparrow < 0$, indicating that the higher excited state $\ket{0\uparrow}$ relaxes to the ground state $\ket{1}$ in this regime at a rate much faster than the relaxation of the lower excited state $\ket{0\downarrow}$.

\subsection{Tunneling rates}

We assume that the tunneling process is spin-conserving and characterized by a spin- and momentum-independent tunneling amplitude $t$. The density of states of the substrate is denoted $\rho_\sigma(\omega)$, that may differ for the two spin species $\sigma=\{\uparrow,\downarrow\}$. The chemical potentials for different spin species align and are denoted by $\mu_s$ for the substrate and $\mu_t$ for the tip, respectively. The occupation of the electronic states of the substrate is given by the Fermi-Dirac distribution $n_{F}(\omega)$.

In the experiment, the magnetic field is sufficiently large such that the substrate is in the normal-conducting state. This provides a simplification, since the density of states is approximately flat at the scale of the Zeeman splitting $E_Z$, and therefore becomes spin-independent. However, we will provide expressions for the rate equations that can account for a future experimental situation where the substrate density of states could be potentially spin-dependent.

\begin{figure}
    \includegraphics[width=\columnwidth]{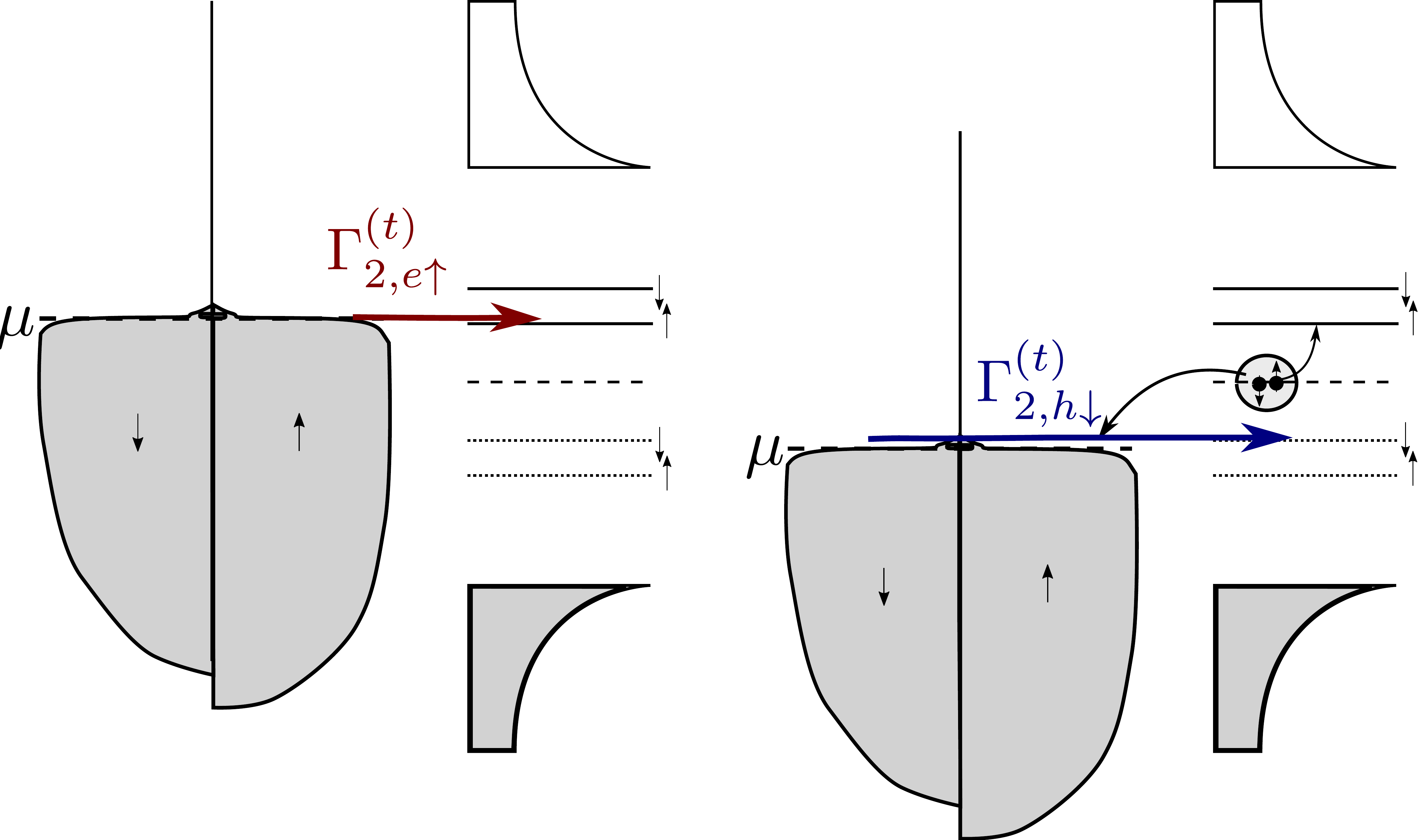}
    \caption{\label{fig:transport_rates} Illustration of two transport processes that add a spin $\uparrow$ quasiparticle to the YSR state, one realized by transporting a spin $\uparrow$ electron, the other realized by transporting a spin $\downarrow$ hole. The latter is equivalent to an electron with spin $\downarrow$ resulting from a Cooper pair splitting event, traveling in the opposite direction of the hole.}
\end{figure}

The rates describing tunneling processes contribute to the total rates $\Gamma_{1\sigma}$ and $\Gamma_{2\sigma}$, as follows. When the initial state is either one of the doublets, $\ket{0\sigma}$, a tunneling process will add a quasiparticle with spin $\bar{\sigma}$, opposite $\sigma$, resulting in the singlet. We distinguish two possibilities: either i. an electron with spin $\bar{\sigma}$ tunnels into the YSR state, with rate denoted by $\Gamma^{(t)}_{2,e\bar{\sigma}}$; or ii. a hole with spin $\sigma$ tunnels into the YSR state, with rate denoted by $\Gamma^{(t)}_{2,h\sigma}$ (see also Fig.~\ref{fig:transport_rates}). Since both processes create a quasiparticle excitation with spin $\bar{\sigma}$ in the YSR state, they add up to give the contribution to $\Gamma_{2\bar{\sigma}}$ due to tunneling, denoted $\Gamma^{(t)}_{2\bar{\sigma}}$.
\begin{align*}
\Gamma^{(t)}_{2\bar{\sigma}} =& \Gamma^{(t)}_{2,e\bar{\sigma}}+\Gamma^{(t)}_{2,h\sigma}.\\
\Gamma^{(t)}_{2,e\bar{\sigma}} =& 2\pi |u|^2 |t|^2 \rho_{\bar{\sigma}}(eV+\epsilon_{\bar{\sigma}})n_{F}(eV+\epsilon_{\bar{\sigma}}).\\
\Gamma^{(t)}_{2,h\sigma} =& 2\pi |v|^2 |t|^2 \rho_{\sigma}(eV-\epsilon_{\bar{\sigma}})\bar{n}_{F}(eV-\epsilon_{\bar{\sigma}}).
\end{align*}
In the expressions above, we have introduced the coherence factors of the YSR state, $u$ and $v$. We denoted by $|u|^2$ the probability to add an electron to the YSR state, while $|v|^2$ represents the probability to add a hole, respectively. Furthermore, we have used the convention that $eV = \mu_t-\mu_s$, where $\mu_s$ and $\mu_t$ are the chemical potentials of the substrate and tip, respectively. We have also introduced the common notation $\bar{n}_F(\omega)=1-n_F(\omega)$ to denote the occupation for holes.

Similarly, we obtain the rates of tunneling processes that remove a quasiparticle from the YSR state, leading to the transition from state $\ket{1}$ to one of the dublet states $\ket{0\sigma}$. A quasiparticle can be removed either by removing from the YSR state an electron with spin $\bar{\sigma}$, $\Gamma^{(t)}_{1,e\bar{\sigma}}$, or a hole with spin $\sigma$, $\Gamma^{(t)}_{1,h\sigma}$. We find in  analogy to the results for adding a quasiparticle above,
\begin{align*}
    \Gamma^{(t)}_{1\bar{\sigma}} =& \Gamma^{(t)}_{1,e\bar{\sigma}}+\Gamma^{(t)}_{1,h\sigma}.\\
    \Gamma^{(t)}_{1,e\bar{\sigma}} =& 2\pi |u|^2 |t|^2 \rho_{\bar{\sigma}}(eV+\epsilon_{\bar{\sigma}})\bar{n}_{F}(eV+\epsilon_{\bar{\sigma}}).\\
    \Gamma^{(t)}_{1,h\sigma} =& 2\pi |v|^2 |t|^2 \rho_{\sigma}(eV-\epsilon_{\bar{\sigma}})n_{F}(eV-\epsilon_{\bar{\sigma}}).
\end{align*}

\subsection{Steady state current}

The electrical current is expressed in terms of tunneling rates and the steady state probabilities. The latter are given by Eq.~(\ref{eq:probabilities}), with the total rates $\Gamma_{1\sigma} = \Gamma^{(i)}_1+\Gamma^{(t)}_{1\sigma}$ and $\Gamma_{2\sigma}=\Gamma^{(i)}_{2\sigma}+\Gamma^{(t)}_{2\sigma}$, given by sum of intrinsic and tunneling rates.

The total steady state current is given by the expression,
\begin{align}
I = e\left[\displaystyle\sum_{\sigma}\left(\Gamma^{(t)}_{2,e\bar{\sigma}}-\Gamma^{(t)}_{2,h\sigma}\right)P_{0\sigma}+\left(\Gamma^{(t)}_{1,h\sigma}-\Gamma^{(t)}_{1,e\bar{\sigma}}\right)P_1\right].
\end{align}
The expression accounts for all charge transfer processes across the tip-substrate junction. The first term accounts for the possibility to add a quasiparticle to the YSR state $\ket{0\sigma}$ either by transporting an electron with spin $\bar{\sigma}$, or a hole with spin $\sigma$. The second term, similarly, accounts for the possibility to remove a quasiparticle with spin $\sigma=\uparrow$ or $\downarrow$ from the YSR state $\ket{1}$, by transporting an electron with spin $\bar{\sigma}$, or a hole with spin $\sigma$.

\begin{figure}
    \includegraphics[width=\columnwidth]{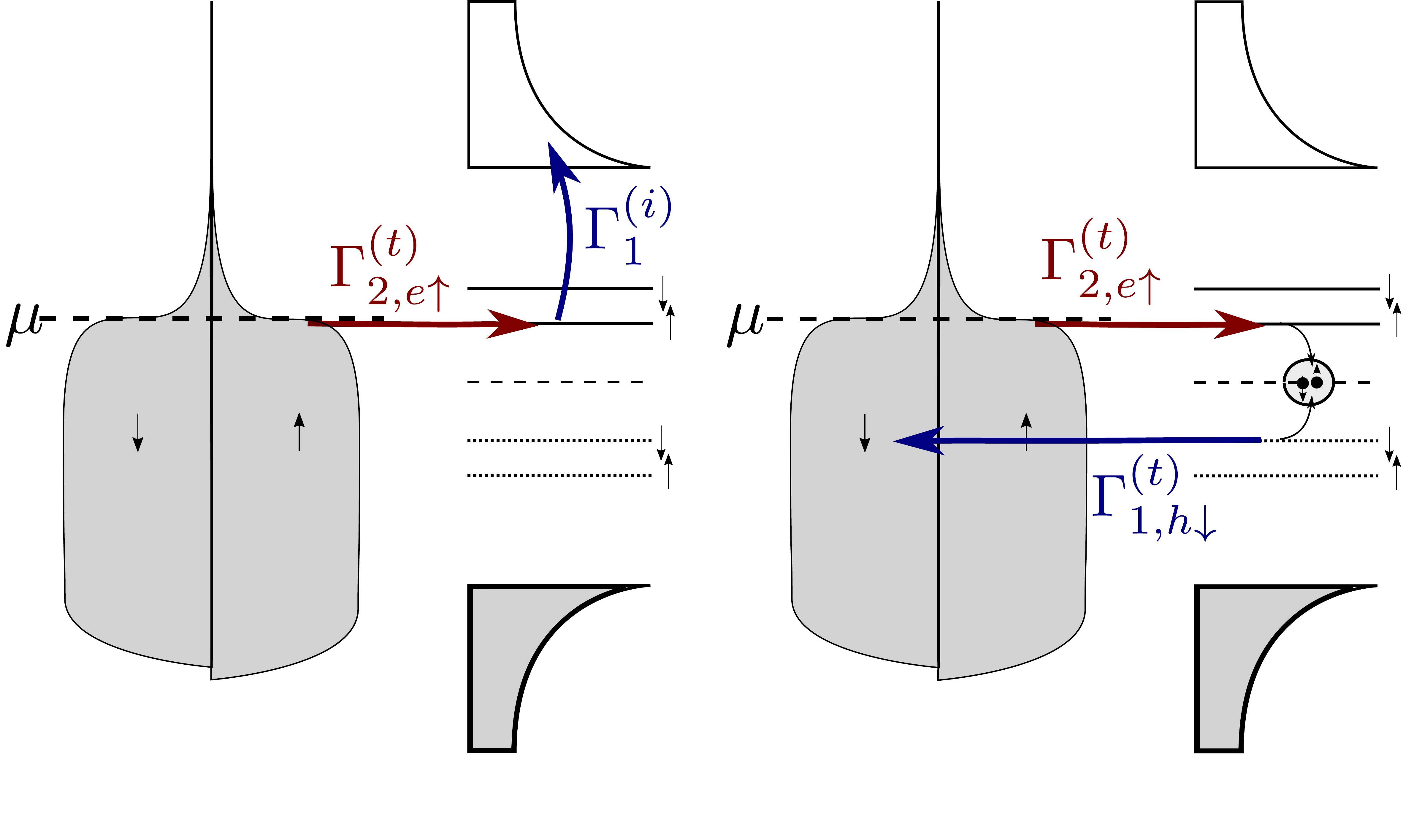}
    \caption{\label{fig:e_and_2e} Two types of elementary transport processes. Left: illustration of an elementary transport process carrying charge $e$. In the first step, a tunneling event leads to the transition $\ket{0\downarrow}$ to $\ket{1}$. In a second step, the state $\ket{1}$ relaxes back to state $\ket{0\downarrow}$ by an intrinsic process without charge transport. Right: illustration of an elementary transport process carrying charge $2e$. The first step is identical to the left illustration. In the second step, a hole tunnels out of the YSR state, restoring the state $\ket{0\downarrow}$ and leading to a total $2e$ charge transfer. The hole tunneling can be seen as an electron tunneling in the opposite direction, forming a Cooper pair with the electron that tunneled in the first step. }
\end{figure}

The total current can be further understood in terms of elementary transport processes. An elementary transport process consists of two transitions: the first changes the occupation of the YSR state, and the second restores the original occupation, thereby completing the transport cycle. We distinguish two types of elementary transport processes: i. when one transition occurs due to a tunneling process, while the other transition is intrinsic, such that a total charge $e$ is transported; and ii. when both transitions occur due to a tunneling process, such that a total charge $2e$ is transported by sequential charge $e$ tunneling events. Fig.~\ref{fig:e_and_2e} illustrates an example of the two types of processes.

\sidecaptionvpos{figure}{c}
\begin{SCfigure*}
    \includegraphics[width=0.78\textwidth]{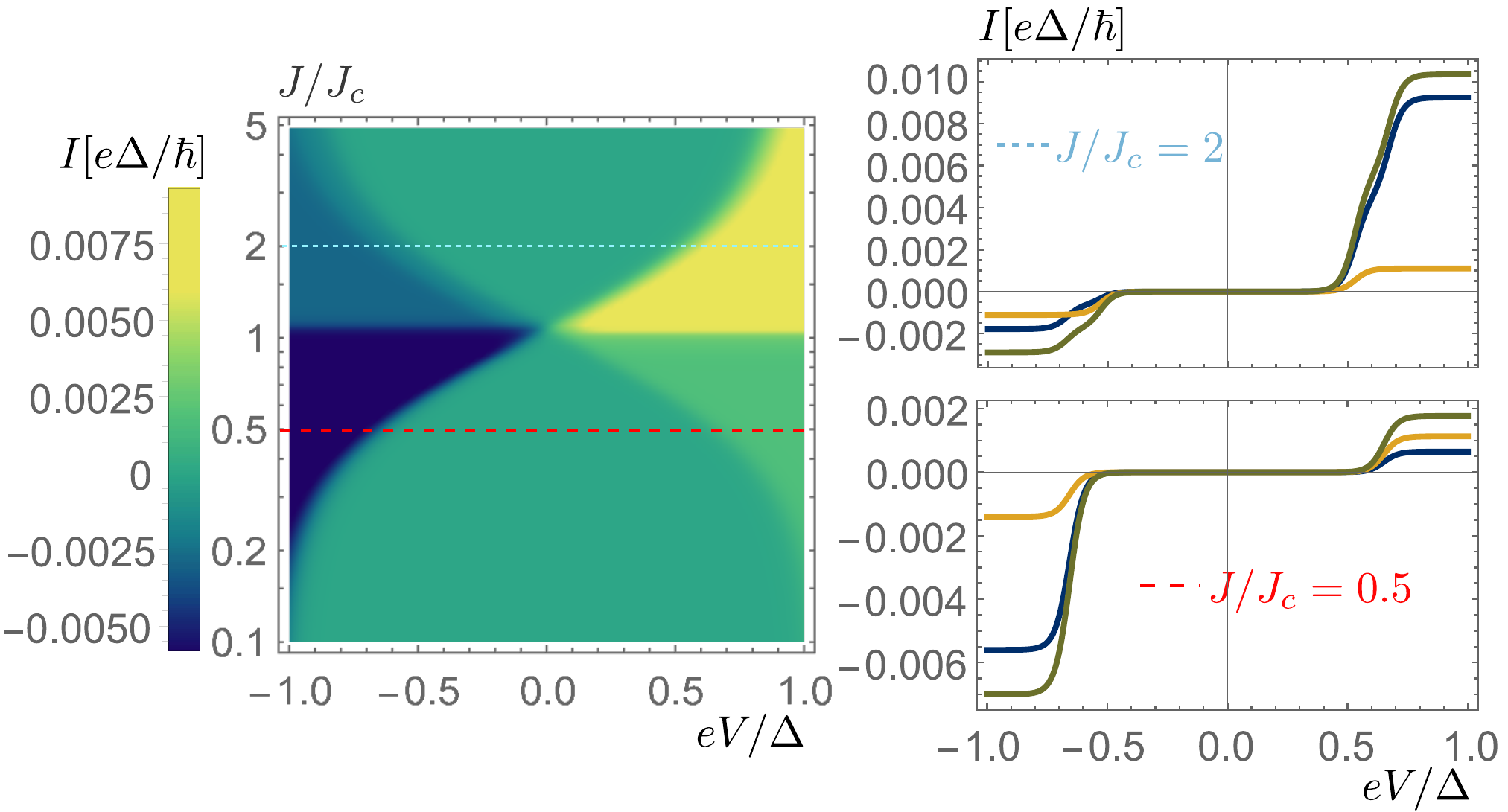}  
    \caption{\label{fig:IV} Current-voltage curves. Left: Density plot of the current as a function of voltage $V$ and exchange coupling $J$. Right: cuts showing the I-V curves before and after the phase transition. The blue and gold curves show the contributions $I_e$ and $I_{2e}$, respectively, while the green curves show the total current. }
\end{SCfigure*}

\sidecaptionvpos{figure}{c}
\begin{SCfigure*}
    \includegraphics[width=0.78\textwidth]{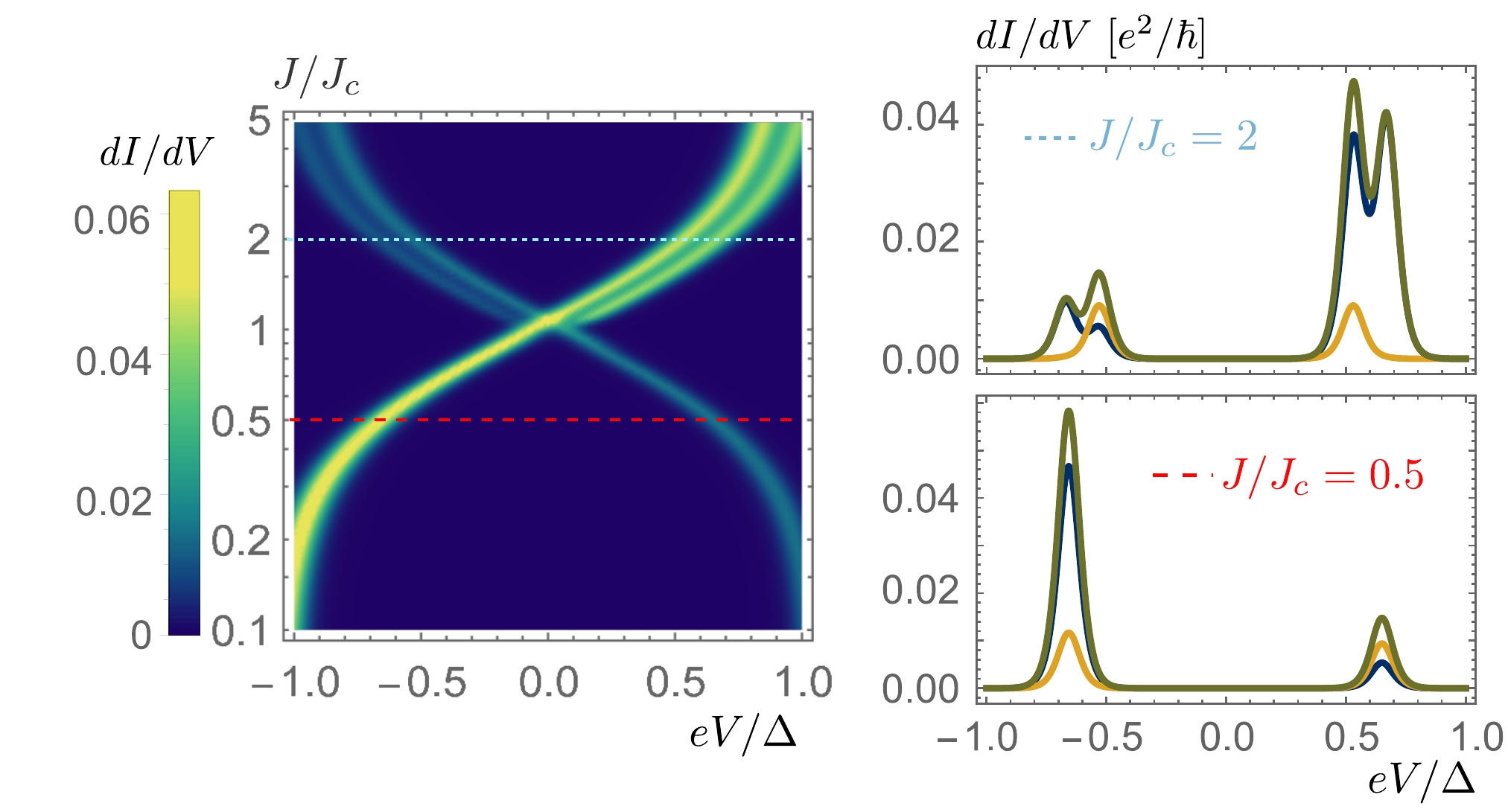}
    \caption{\label{fig:dIdV} Differential conductance curves. Left: Density plot of the differential conductance as a function of voltage $V$ and exchange coupling $J$. Right: cuts showing the differential conductance curves before and after the phase transition. The blue and gold curves show the contributions $dI_e/dV$ and $dI_{2e}/dV$, respectively, while the green curves show the differential conductance. }
\end{SCfigure*}

The total steady state current is the sum of currents contributed by the two types of processes,
\[I = I_e + I_{2e}.\]

\subsubsection{Charge $e$ elementary transport process}

For this transport process, charge is transported in a single tunneling event. We must account both for forward transport and for backward transport, as follows.
\begin{align}
I_e = e\displaystyle\sum_{\sigma\sigma'}\frac{\left(\Gamma^{(t)}_{2,e\bar{\sigma}}-\Gamma^{(t)}_{2,h\sigma}\right)\Gamma_1^{(i)} + \Gamma_{2\bar{\sigma}}^{(i)}\left(\Gamma_{1,h\sigma'}^{(t)}-\Gamma^{(t)}_{1,e\bar{\sigma}'}\right)}{(\Gamma_{1\uparrow}+\Gamma_{1\downarrow})}P_{0\sigma}.
\end{align}
The first term, proportional to $P_{0\sigma}$, describes the process when an electron, or a hole, tunnels into the YSR state changing the occupation from $\ket{0\sigma}$ to $\ket{1}$, as shown in Fig.~\ref{fig:transport_rates}. The second step of the elementary process occurs therefore via an intrinsic process, restoring an unoccupied YSR state $\ket{0\sigma'}$, without charge transport. The total process is depicted in the left side of Fig.~\ref{fig:e_and_2e}. Alternatively, the transition from $\ket{0\sigma}$ to $\ket{1}$ could occur by an intrinsic process, while the second step, from $\ket{1}$ to $\ket{0\sigma'}$, could occur by tunneling.

Note that the information about the energy of the states and filling factors of the substrate are all encoded in the rates and indirectly, in the steady state probabilities. Therefore, the expressions apply for all regimes, free spin, intermediate, as well as the screened spin regime.

\subsubsection{Charge $2e$ elementary transport process}

When both steps of the elementary transport process involve a tunneling event, such as the process depicted in the right side of Fig.~\ref{fig:e_and_2e}, the total transported charge is $2e$. The transport of $2e$ charge is reminiscent of Andreev reflection. Indeed, the two charged particles involved in transport change the number of Cooper pairs in the tip condensate by one. However, there is also an important difference, the process described here is a sequential process consisting of two single particle transport events.

Elementary transport processes involving two sequential tunneling events have the following structure, \[\ket{0\sigma}\xrightarrow{\Gamma^{(t)}_{2,e\bar{\sigma}};\Gamma^{(t)}_{2,h\sigma}}\ket{1}\xrightarrow{\Gamma^{(t)}_{1,h\sigma'};\Gamma^{(t)}_{1,e\bar{\sigma}'}}\ket{0\sigma'}.\]
The combination of $\Gamma^{(t)}_{2,e\bar{\sigma}}$ and $\Gamma^{(t)}_{1,e\bar{\sigma}'}$ describes an electron tunneling back and forth across the junction, giving rise to current noise, but without net charge transport. Similarly, the combination of $\Gamma^{(t)}_{2,h\sigma}$ and $\Gamma^{(t)}_{1,h\sigma'}$ describes tunneling back and forth of a hole.

The transport of $2e$ charge arises from combining tunneling of an electron into the tip $\Gamma^{(t)}_{2,e\bar{\sigma}}$
with the subsequent tunneling of a hole out of the tip $\Gamma^{(t)}_{1,h\sigma'}$. 
Note that, as a consequence of the assumption in our model that tunneling can flip the spin of the impurity, all combinations $\{\sigma, \sigma'\}$ are possible for the spin of the tunneling electron and hole.

While the process above transports a charge $2e$ from the substrate to the tip, the reverse process is obtained by combining $\Gamma^{(t)}_{2,h\sigma}$ with $\Gamma^{(t)}_{1,e\bar{\sigma}'}$, which transports a charge $2e$ from the tip to the substrate. 

The total resulting current is given by the sum over all possible $2e$ processes
\begin{align}
I_{2e} = 2e\displaystyle\sum_{\sigma\sigma'}\left(\frac{\Gamma^{(t)}_{1,h\sigma'}\Gamma^{(t)}_{2,e\bar{\sigma}}}{(\Gamma_{1\uparrow}+\Gamma_{1\downarrow})}-\frac{\Gamma^{(t)}_{2,h\sigma}\Gamma^{(t)}_{1,e\bar{\sigma}'}}{(\Gamma_{1\uparrow}+\Gamma_{1\downarrow})}\right)P_{0\sigma}.
\end{align}
We note that typically, only a few of the terms contribute significantly to the total current.

\section{results}
\label{sec:results}

The experimentally relevant regime is defined by $\Delta\simeq 0.6$ $meV$, $E_Z\simeq 0.07 \Delta$ corresponding to $B=750$ mT and a g-factor of $g=2$, and $k_BT\simeq 0.014 \Delta$. In this regime, the results reproduce the steps in current that arise when the voltage aligns the chemical potential of the substrate with an electronic transition of the tip YSR state (see Fig.~\ref{fig:IV}). These current steps manifest as peaks in the differential conductance, as seen in Fig.~\ref{fig:dIdV}.

In the free spin regime, a single peak is found in the differential conductance, corresponding to the electronic transition from the ground state $\ket{0\downarrow}$ to the singlet state $\ket{1}$. Transport processes starting in the higher energy spin state $\ket{0\uparrow}$ are thermally suppressed. In contrast, in the screened spin regime, two peaks are seen, contributed by transitions from the ground state $\ket{1}$ to either excited states $\ket{0\downarrow}$ or $\ket{0\uparrow}$. The distance between the two peaks is directly given by twice the Zeeman energy.

Both charge $e$ and charge $2e$ elementary processes contribute significantly in the experimentally relevant parameter regime, although typically charge $e$ processes dominate.

The theoretical result presented in Fig. 3 of the main text, accounts for the fact that in the experiment, approaching the tip increases the transmission of the tunnel barrier, $\tau\equiv |t|^2$, and also modifies the effective exchange coupling $J$, that leads to the modification of the YSR state excitation energies, $\epsilon_\uparrow$ and $\epsilon_\downarrow$. We have used the simplest method to account for the simultaneous dependencies, by assuming a linear dependence of the YSR excitation energies on the transmission $\tau$,
\begin{align}
\epsilon_\sigma(\tau) = \alpha \tau +\epsilon_0 +\sigma E_Z/2.
\end{align}
With this linear dependence, the result of the rate equation model reproduces well the measured data. The parameters $\alpha$ and $\epsilon_0$ can be fitted independently, using the position of YSR peaks in absence of magnetic field. The only free parameters that concern electronic transport in presence of magnetic field are related to the intrinsic relaxation rate $\Gamma_R/\Delta=0.01$, the tunneling rate $\Gamma_t/\Delta=2\pi|t|^2\rho/\Delta=0.01$, and the asymmetry in the coherence factors of the YSR state, $|u|^2/|v|^2=7$. These are fitting parameters for the plot in Fig. 3 in the main text.


\begin{thebibliography}{52}%
\makeatletter
\providecommand \@ifxundefined [1]{%
 \@ifx{#1\undefined}
}%
\providecommand \@ifnum [1]{%
 \ifnum #1\expandafter \@firstoftwo
 \else \expandafter \@secondoftwo
 \fi
}%
\providecommand \@ifx [1]{%
 \ifx #1\expandafter \@firstoftwo
 \else \expandafter \@secondoftwo
 \fi
}%
\providecommand \natexlab [1]{#1}%
\providecommand \emph  [1]{``#1''}%
\providecommand \bibnamefont  [1]{#1}%
\providecommand \bibfnamefont [1]{#1}%
\providecommand \citenamefont [1]{#1}%
\providecommand \href@noop [0]{\@secondoftwo}%
\providecommand \href [0]{\begingroup \@sanitize@url \@href}%
\providecommand \@href[1]{\@@startlink{#1}\@@href}%
\providecommand \@@href[1]{\endgroup#1\@@endlink}%
\providecommand \@sanitize@url [0]{\catcode `\\12\catcode `\$12\catcode
  `\&12\catcode `\#12\catcode `\^12\catcode `\_12\catcode `\%12\relax}%
\providecommand \@@startlink[1]{}%
\providecommand \@@endlink[0]{}%
\providecommand \url  [0]{\begingroup\@sanitize@url \@url }%
\providecommand \@url [1]{\endgroup\@href {#1}{\urlprefix }}%
\providecommand \urlprefix  [0]{URL }%
\providecommand \Eprint [0]{\href }%
\providecommand \doibase [0]{http://dx.doi.org/}%
\providecommand \selectlanguage [0]{\@gobble}%
\providecommand \bibinfo  [0]{\@secondoftwo}%
\providecommand \bibfield  [0]{\@secondoftwo}%
\providecommand \translation [1]{[#1]}%
\providecommand \BibitemOpen [0]{}%
\providecommand \bibitemStop [0]{}%
\providecommand \bibitemNoStop [0]{.\EOS\space}%
\providecommand \EOS [0]{\spacefactor3000\relax}%
\providecommand \BibitemShut  [1]{\csname bibitem#1\endcsname}%
\let\auto@bib@innerbib\@empty
%</preamble>
\bibitem [{\citenamefont {Yu}(1965)}]{yu_bound_1965}%
  \BibitemOpen
  \bibfield  {author} {\bibinfo {author} {\bibfnamefont {L.}~\bibnamefont
  {Yu}},\ }\bibfield  {title} {{\selectlanguage {english}\emph {\bibinfo
  {title} {Bound state in superconductors with paramagnetic impurities},}\
  }}\href {http://wulixb.iphy.ac.cn/EN/abstract/abstract851.shtml} {\bibfield
  {journal} {\bibinfo  {journal} {Acta Phys. Sin.}\ }\textbf {\bibinfo {volume}
  {21}},\ \bibinfo {pages} {75} (\bibinfo {year} {1965})}\BibitemShut {NoStop}%
\bibitem [{\citenamefont {Shiba}(1968)}]{shiba_classical_1968}%
  \BibitemOpen
  \bibfield  {author} {\bibinfo {author} {\bibfnamefont {H.}~\bibnamefont
  {Shiba}},\ }\bibfield  {title} {{\selectlanguage {english}\emph {\bibinfo
  {title} {Classical spins in superconductors},}\ }}\href {\doibase
  10.1143/PTP.40.435} {\bibfield  {journal} {\bibinfo  {journal} {Prog. Theor.
  Phys.}\ }\textbf {\bibinfo {volume} {40}},\ \bibinfo {pages} {435} (\bibinfo
  {year} {1968})}\BibitemShut {NoStop}%
\bibitem [{\citenamefont {Rusinov}(1969)}]{rusinov_superconductivity_1969}%
  \BibitemOpen
  \bibfield  {author} {\bibinfo {author} {\bibfnamefont {A.~I.}\ \bibnamefont
  {Rusinov}},\ }\bibfield  {title} {\emph {\bibinfo {title} {Superconductivity
  near a paramagmetic impurity},}\ }\href
  {http://jetpletters.ru/ps/1658/article_25295.shtml} {\bibfield  {journal}
  {\bibinfo  {journal} {JETP Letters}\ }\textbf {\bibinfo {volume} {9}},\
  \bibinfo {pages} {85} (\bibinfo {year} {1969})}\BibitemShut {NoStop}%
\bibitem [{\citenamefont {Yazdani}\ \emph {et~al.}(1997)\citenamefont
  {Yazdani}, \citenamefont {Jones}, \citenamefont {Lutz}, \citenamefont
  {Crommie},\ and\ \citenamefont {Eigler}}]{yazdani_probing_1997}%
  \BibitemOpen
  \bibfield  {author} {\bibinfo {author} {\bibfnamefont {A.}~\bibnamefont
  {Yazdani}}, \bibinfo {author} {\bibfnamefont {B.~A.}\ \bibnamefont {Jones}},
  \bibinfo {author} {\bibfnamefont {C.~P.}\ \bibnamefont {Lutz}}, \bibinfo
  {author} {\bibfnamefont {M.~F.}\ \bibnamefont {Crommie}}, \ and\ \bibinfo
  {author} {\bibfnamefont {D.~M.}\ \bibnamefont {Eigler}},\ }\bibfield  {title}
  {\emph {\bibinfo {title} {Probing the {Local} {Effects} of {Magnetic}
  {Impurities} on {Superconductivity}},}\ }\href {\doibase
  10.1126/science.275.5307.1767} {\bibfield  {journal} {\bibinfo  {journal}
  {Science}\ }\textbf {\bibinfo {volume} {275}},\ \bibinfo {pages} {1767}
  (\bibinfo {year} {1997})}\BibitemShut {NoStop}%
\bibitem [{\citenamefont {Balatsky}\ \emph {et~al.}(2006)\citenamefont
  {Balatsky}, \citenamefont {Vekhter},\ and\ \citenamefont
  {Zhu}}]{balatsky_impurity-induced_2006}%
  \BibitemOpen
  \bibfield  {author} {\bibinfo {author} {\bibfnamefont {A.~V.}\ \bibnamefont
  {Balatsky}}, \bibinfo {author} {\bibfnamefont {I.}~\bibnamefont {Vekhter}}, \
  and\ \bibinfo {author} {\bibfnamefont {J.-X.}\ \bibnamefont {Zhu}},\
  }\bibfield  {title} {\emph {\bibinfo {title} {Impurity-induced states in
  conventional and unconventional superconductors},}\ }\href {\doibase
  10.1103/RevModPhys.78.373} {\bibfield  {journal} {\bibinfo  {journal}
  {Reviews of Modern Physics}\ }\textbf {\bibinfo {volume} {78}},\ \bibinfo
  {pages} {373} (\bibinfo {year} {2006})}\BibitemShut {NoStop}%
\bibitem [{\citenamefont {Heinrich}\ \emph {et~al.}(2018)\citenamefont
  {Heinrich}, \citenamefont {Pascual},\ and\ \citenamefont
  {Franke}}]{heinrich_single_2018}%
  \BibitemOpen
  \bibfield  {author} {\bibinfo {author} {\bibfnamefont {B.~W.}\ \bibnamefont
  {Heinrich}}, \bibinfo {author} {\bibfnamefont {J.~I.}\ \bibnamefont
  {Pascual}}, \ and\ \bibinfo {author} {\bibfnamefont {K.~J.}\ \bibnamefont
  {Franke}},\ }\bibfield  {title} {\emph {\bibinfo {title} {Single magnetic
  adsorbates on s-wave superconductors},}\ }\href
  {http://arxiv.org/abs/1705.03672} {\bibfield  {journal} {\bibinfo  {journal}
  {Prog. Surf. Sci.}\ }\textbf {\bibinfo {volume} {93}},\ \bibinfo {pages} {1}
  (\bibinfo {year} {2018})}\BibitemShut {NoStop}%
\bibitem [{\citenamefont {Farinacci}\ \emph {et~al.}(2018)\citenamefont
  {Farinacci}, \citenamefont {Ahmadi}, \citenamefont {Reecht}, \citenamefont
  {Ruby}, \citenamefont {Bogdanoff}, \citenamefont {Peters}, \citenamefont
  {Heinrich}, \citenamefont {von Oppen},\ and\ \citenamefont
  {Franke}}]{farinacci_tuning_2018}%
  \BibitemOpen
  \bibfield  {author} {\bibinfo {author} {\bibfnamefont {L.}~\bibnamefont
  {Farinacci}}, \bibinfo {author} {\bibfnamefont {G.}~\bibnamefont {Ahmadi}},
  \bibinfo {author} {\bibfnamefont {G.}~\bibnamefont {Reecht}}, \bibinfo
  {author} {\bibfnamefont {M.}~\bibnamefont {Ruby}}, \bibinfo {author}
  {\bibfnamefont {N.}~\bibnamefont {Bogdanoff}}, \bibinfo {author}
  {\bibfnamefont {O.}~\bibnamefont {Peters}}, \bibinfo {author} {\bibfnamefont
  {B.~W.}\ \bibnamefont {Heinrich}}, \bibinfo {author} {\bibfnamefont
  {F.}~\bibnamefont {von Oppen}}, \ and\ \bibinfo {author} {\bibfnamefont
  {K.~J.}\ \bibnamefont {Franke}},\ }\bibfield  {title} {\emph {\bibinfo
  {title} {Tuning the {Coupling} of an {Individual} {Magnetic} {Impurity} to a
  {Superconductor}: {Quantum} {Phase} {Transition} and {Transport}},}\ }\href
  {\doibase 10.1103/PhysRevLett.121.196803} {\bibfield  {journal} {\bibinfo
  {journal} {Physical Review Letters}\ }\textbf {\bibinfo {volume} {121}},\
  \bibinfo {pages} {196803} (\bibinfo {year} {2018})}\BibitemShut {NoStop}%
\bibitem [{\citenamefont {Küster}\ \emph {et~al.}(2021)\citenamefont
  {Küster}, \citenamefont {Brinker}, \citenamefont {Lounis}, \citenamefont
  {Parkin},\ and\ \citenamefont {Sessi}}]{kuster_long_2021}%
  \BibitemOpen
  \bibfield  {author} {\bibinfo {author} {\bibfnamefont {F.}~\bibnamefont
  {Küster}}, \bibinfo {author} {\bibfnamefont {S.}~\bibnamefont {Brinker}},
  \bibinfo {author} {\bibfnamefont {S.}~\bibnamefont {Lounis}}, \bibinfo
  {author} {\bibfnamefont {S.~S.~P.}\ \bibnamefont {Parkin}}, \ and\ \bibinfo
  {author} {\bibfnamefont {P.}~\bibnamefont {Sessi}},\ }\bibfield  {title}
  {{\selectlanguage {english}\emph {\bibinfo {title} {Long range and highly
  tunable interaction between local spins coupled to a superconducting
  condensate},}\ }}\href {\doibase 10.1038/s41467-021-26802-x} {\bibfield
  {journal} {\bibinfo  {journal} {Nature Communications}\ }\textbf {\bibinfo
  {volume} {12}},\ \bibinfo {pages} {6722} (\bibinfo {year}
  {2021})}\BibitemShut {NoStop}%
\bibitem [{\citenamefont {Franke}\ \emph {et~al.}(2011)\citenamefont {Franke},
  \citenamefont {Schulze},\ and\ \citenamefont
  {Pascual}}]{franke_competition_2011}%
  \BibitemOpen
  \bibfield  {author} {\bibinfo {author} {\bibfnamefont {K.~J.}\ \bibnamefont
  {Franke}}, \bibinfo {author} {\bibfnamefont {G.}~\bibnamefont {Schulze}}, \
  and\ \bibinfo {author} {\bibfnamefont {J.~I.}\ \bibnamefont {Pascual}},\
  }\bibfield  {title} {\emph {\bibinfo {title} {Competition of
  {Superconducting} {Phenomena} and {Kondo} {Screening} at the {Nanoscale}},}\
  }\href {\doibase 10.1126/science.1202204} {\bibfield  {journal} {\bibinfo
  {journal} {Science}\ }\textbf {\bibinfo {volume} {332}},\ \bibinfo {pages}
  {940} (\bibinfo {year} {2011})}\BibitemShut {NoStop}%
\bibitem [{\citenamefont {Kamlapure}\ \emph {et~al.}(2021)\citenamefont
  {Kamlapure}, \citenamefont {Cornils}, \citenamefont {Žitko}, \citenamefont
  {Valentyuk}, \citenamefont {Mozara}, \citenamefont {Pradhan}, \citenamefont
  {Fransson}, \citenamefont {Lichtenstein}, \citenamefont {Wiebe},\ and\
  \citenamefont {Wiesendanger}}]{kamlapure_correlation_2021}%
  \BibitemOpen
  \bibfield  {author} {\bibinfo {author} {\bibfnamefont {A.}~\bibnamefont
  {Kamlapure}}, \bibinfo {author} {\bibfnamefont {L.}~\bibnamefont {Cornils}},
  \bibinfo {author} {\bibfnamefont {R.}~\bibnamefont {Žitko}}, \bibinfo
  {author} {\bibfnamefont {M.}~\bibnamefont {Valentyuk}}, \bibinfo {author}
  {\bibfnamefont {R.}~\bibnamefont {Mozara}}, \bibinfo {author} {\bibfnamefont
  {S.}~\bibnamefont {Pradhan}}, \bibinfo {author} {\bibfnamefont
  {J.}~\bibnamefont {Fransson}}, \bibinfo {author} {\bibfnamefont {A.~I.}\
  \bibnamefont {Lichtenstein}}, \bibinfo {author} {\bibfnamefont
  {J.}~\bibnamefont {Wiebe}}, \ and\ \bibinfo {author} {\bibfnamefont
  {R.}~\bibnamefont {Wiesendanger}},\ }\bibfield  {title} {\emph {\bibinfo
  {title} {Correlation of {Yu}–{Shiba}–{Rusinov} {States} and {Kondo}
  {Resonances} in {Artificial} {Spin} {Arrays} on an s-{Wave}
  {Superconductor}},}\ }\href {\doibase 10.1021/acs.nanolett.1c00387}
  {\bibfield  {journal} {\bibinfo  {journal} {Nano Letters}\ }\textbf {\bibinfo
  {volume} {21}},\ \bibinfo {pages} {6748} (\bibinfo {year}
  {2021})}\BibitemShut {NoStop}%
\bibitem [{\citenamefont {Malavolti}\ \emph {et~al.}(2018)\citenamefont
  {Malavolti}, \citenamefont {Briganti}, \citenamefont {Hänze}, \citenamefont
  {Serrano}, \citenamefont {Cimatti}, \citenamefont {McMurtrie}, \citenamefont
  {Otero}, \citenamefont {Ohresser}, \citenamefont {Totti}, \citenamefont
  {Mannini}, \citenamefont {Sessoli},\ and\ \citenamefont
  {Loth}}]{malavolti_tunable_2018}%
  \BibitemOpen
  \bibfield  {author} {\bibinfo {author} {\bibfnamefont {L.}~\bibnamefont
  {Malavolti}}, \bibinfo {author} {\bibfnamefont {M.}~\bibnamefont {Briganti}},
  \bibinfo {author} {\bibfnamefont {M.}~\bibnamefont {Hänze}}, \bibinfo
  {author} {\bibfnamefont {G.}~\bibnamefont {Serrano}}, \bibinfo {author}
  {\bibfnamefont {I.}~\bibnamefont {Cimatti}}, \bibinfo {author} {\bibfnamefont
  {G.}~\bibnamefont {McMurtrie}}, \bibinfo {author} {\bibfnamefont
  {E.}~\bibnamefont {Otero}}, \bibinfo {author} {\bibfnamefont
  {P.}~\bibnamefont {Ohresser}}, \bibinfo {author} {\bibfnamefont
  {F.}~\bibnamefont {Totti}}, \bibinfo {author} {\bibfnamefont
  {M.}~\bibnamefont {Mannini}}, \bibinfo {author} {\bibfnamefont
  {R.}~\bibnamefont {Sessoli}}, \ and\ \bibinfo {author} {\bibfnamefont
  {S.}~\bibnamefont {Loth}},\ }\bibfield  {title} {\emph {\bibinfo {title}
  {Tunable {Spin}–{Superconductor} {Coupling} of {Spin} 1/2 {Vanadyl}
  {Phthalocyanine} {Molecules}},}\ }\href {\doibase
  10.1021/acs.nanolett.8b03921} {\bibfield  {journal} {\bibinfo  {journal}
  {Nano Letters}\ }\textbf {\bibinfo {volume} {18}},\ \bibinfo {pages} {7955}
  (\bibinfo {year} {2018})}\BibitemShut {NoStop}%
\bibitem [{\citenamefont {Bauer}\ \emph {et~al.}(2013)\citenamefont {Bauer},
  \citenamefont {Pascual},\ and\ \citenamefont
  {Franke}}]{bauer_microscopic_2013}%
  \BibitemOpen
  \bibfield  {author} {\bibinfo {author} {\bibfnamefont {J.}~\bibnamefont
  {Bauer}}, \bibinfo {author} {\bibfnamefont {J.~I.}\ \bibnamefont {Pascual}},
  \ and\ \bibinfo {author} {\bibfnamefont {K.~J.}\ \bibnamefont {Franke}},\
  }\bibfield  {title} {\emph {\bibinfo {title} {Microscopic resolution of the
  interplay of {Kondo} screening and superconducting pairing:
  {Mn}-phthalocyanine molecules adsorbed on superconducting {Pb}(111)},}\
  }\href {\doibase 10.1103/PhysRevB.87.075125} {\bibfield  {journal} {\bibinfo
  {journal} {Physical Review B}\ }\textbf {\bibinfo {volume} {87}},\ \bibinfo
  {pages} {075125} (\bibinfo {year} {2013})}\BibitemShut {NoStop}%
\bibitem [{\citenamefont {Chatzopoulos}\ \emph {et~al.}(2021)\citenamefont
  {Chatzopoulos}, \citenamefont {Cho}, \citenamefont {Bastiaans}, \citenamefont
  {Steffensen}, \citenamefont {Bouwmeester}, \citenamefont {Akbari},
  \citenamefont {Gu}, \citenamefont {Paaske}, \citenamefont {Andersen},\ and\
  \citenamefont {Allan}}]{chatzopoulos_spatially_2021}%
  \BibitemOpen
  \bibfield  {author} {\bibinfo {author} {\bibfnamefont {D.}~\bibnamefont
  {Chatzopoulos}}, \bibinfo {author} {\bibfnamefont {D.}~\bibnamefont {Cho}},
  \bibinfo {author} {\bibfnamefont {K.~M.}\ \bibnamefont {Bastiaans}}, \bibinfo
  {author} {\bibfnamefont {G.~O.}\ \bibnamefont {Steffensen}}, \bibinfo
  {author} {\bibfnamefont {D.}~\bibnamefont {Bouwmeester}}, \bibinfo {author}
  {\bibfnamefont {A.}~\bibnamefont {Akbari}}, \bibinfo {author} {\bibfnamefont
  {G.}~\bibnamefont {Gu}}, \bibinfo {author} {\bibfnamefont {J.}~\bibnamefont
  {Paaske}}, \bibinfo {author} {\bibfnamefont {B.~M.}\ \bibnamefont
  {Andersen}}, \ and\ \bibinfo {author} {\bibfnamefont {M.~P.}\ \bibnamefont
  {Allan}},\ }\bibfield  {title} {{\selectlanguage {english}\emph {\bibinfo
  {title} {Spatially dispersing {Yu}-{Shiba}-{Rusinov} states in the
  unconventional superconductor {FeTe0}.{55Se0}.45},}\ }}\href {\doibase
  10.1038/s41467-020-20529-x} {\bibfield  {journal} {\bibinfo  {journal}
  {Nature Communications}\ }\textbf {\bibinfo {volume} {12}},\ \bibinfo {pages}
  {298} (\bibinfo {year} {2021})}\BibitemShut {NoStop}%
\bibitem [{\citenamefont {Ruby}\ \emph {et~al.}(2015)\citenamefont {Ruby},
  \citenamefont {Pientka}, \citenamefont {Peng}, \citenamefont {von Oppen},
  \citenamefont {Heinrich},\ and\ \citenamefont
  {Franke}}]{ruby_tunneling_2015}%
  \BibitemOpen
  \bibfield  {author} {\bibinfo {author} {\bibfnamefont {M.}~\bibnamefont
  {Ruby}}, \bibinfo {author} {\bibfnamefont {F.}~\bibnamefont {Pientka}},
  \bibinfo {author} {\bibfnamefont {Y.}~\bibnamefont {Peng}}, \bibinfo {author}
  {\bibfnamefont {F.}~\bibnamefont {von Oppen}}, \bibinfo {author}
  {\bibfnamefont {B.~W.}\ \bibnamefont {Heinrich}}, \ and\ \bibinfo {author}
  {\bibfnamefont {K.~J.}\ \bibnamefont {Franke}},\ }\bibfield  {title} {\emph
  {\bibinfo {title} {Tunneling {Processes} into {Localized} {Subgap} {States}
  in {Superconductors}},}\ }\href {\doibase 10.1103/PhysRevLett.115.087001}
  {\bibfield  {journal} {\bibinfo  {journal} {Phys. Rev. Lett.}\ }\textbf
  {\bibinfo {volume} {115}},\ \bibinfo {pages} {087001} (\bibinfo {year}
  {2015})}\BibitemShut {NoStop}%
\bibitem [{\citenamefont {Farinacci}\ \emph {et~al.}(2020)\citenamefont
  {Farinacci}, \citenamefont {Ahmadi}, \citenamefont {Ruby}, \citenamefont
  {Reecht}, \citenamefont {Heinrich}, \citenamefont {Czekelius}, \citenamefont
  {von Oppen},\ and\ \citenamefont {Franke}}]{farinacci_interfering_2020}%
  \BibitemOpen
  \bibfield  {author} {\bibinfo {author} {\bibfnamefont {L.}~\bibnamefont
  {Farinacci}}, \bibinfo {author} {\bibfnamefont {G.}~\bibnamefont {Ahmadi}},
  \bibinfo {author} {\bibfnamefont {M.}~\bibnamefont {Ruby}}, \bibinfo {author}
  {\bibfnamefont {G.}~\bibnamefont {Reecht}}, \bibinfo {author} {\bibfnamefont
  {B.~W.}\ \bibnamefont {Heinrich}}, \bibinfo {author} {\bibfnamefont
  {C.}~\bibnamefont {Czekelius}}, \bibinfo {author} {\bibfnamefont
  {F.}~\bibnamefont {von Oppen}}, \ and\ \bibinfo {author} {\bibfnamefont
  {K.~J.}\ \bibnamefont {Franke}},\ }\bibfield  {title} {\emph {\bibinfo
  {title} {Interfering {Tunneling} {Paths} through {Magnetic} {Molecules} on
  {Superconductors}: {Asymmetries} of {Kondo} and {Yu}-{Shiba}-{Rusinov}
  {Resonances}},}\ }\href {\doibase 10.1103/PhysRevLett.125.256805} {\bibfield
  {journal} {\bibinfo  {journal} {Physical Review Letters}\ }\textbf {\bibinfo
  {volume} {125}},\ \bibinfo {pages} {256805} (\bibinfo {year}
  {2020})}\BibitemShut {NoStop}%
\bibitem [{\citenamefont {Brand}\ \emph {et~al.}(2018)\citenamefont {Brand},
  \citenamefont {Gozdzik}, \citenamefont {N{\'e}el}, \citenamefont {Lado},
  \citenamefont {Fern{\'a}ndez-Rossier},\ and\ \citenamefont
  {Kr{\"o}ger}}]{brand_electron_2018}%
  \BibitemOpen
  \bibfield  {author} {\bibinfo {author} {\bibfnamefont {J.}~\bibnamefont
  {Brand}}, \bibinfo {author} {\bibfnamefont {S.}~\bibnamefont {Gozdzik}},
  \bibinfo {author} {\bibfnamefont {N.}~\bibnamefont {N{\'e}el}}, \bibinfo
  {author} {\bibfnamefont {J.~L.}\ \bibnamefont {Lado}}, \bibinfo {author}
  {\bibfnamefont {J.}~\bibnamefont {Fern{\'a}ndez-Rossier}}, \ and\ \bibinfo
  {author} {\bibfnamefont {J.}~\bibnamefont {Kr{\"o}ger}},\ }\bibfield  {title}
  {\emph {\bibinfo {title} {Electron and {Cooper}-pair transport across a
  single magnetic molecule explored with a scanning tunneling microscope},}\
  }\href {\doibase 10.1103/PhysRevB.97.195429} {\bibfield  {journal} {\bibinfo
  {journal} {Physical Review B}\ }\textbf {\bibinfo {volume} {97}},\ \bibinfo
  {pages} {195429} (\bibinfo {year} {2018})}\BibitemShut {NoStop}%
\bibitem [{\citenamefont {Kezilebieke}\ \emph {et~al.}(2019)\citenamefont
  {Kezilebieke}, \citenamefont {Žitko}, \citenamefont {Dvorak}, \citenamefont
  {Ojanen},\ and\ \citenamefont {Liljeroth}}]{kezilebieke_observation_2019}%
  \BibitemOpen
  \bibfield  {author} {\bibinfo {author} {\bibfnamefont {S.}~\bibnamefont
  {Kezilebieke}}, \bibinfo {author} {\bibfnamefont {R.}~\bibnamefont {Žitko}},
  \bibinfo {author} {\bibfnamefont {M.}~\bibnamefont {Dvorak}}, \bibinfo
  {author} {\bibfnamefont {T.}~\bibnamefont {Ojanen}}, \ and\ \bibinfo {author}
  {\bibfnamefont {P.}~\bibnamefont {Liljeroth}},\ }\bibfield  {title} {\emph
  {\bibinfo {title} {Observation of {Coexistence} of {Yu}-{Shiba}-{Rusinov}
  {States} and {Spin}-{Flip} {Excitations}},}\ }\href {\doibase
  10.1021/acs.nanolett.9b01583} {\bibfield  {journal} {\bibinfo  {journal}
  {Nano Letters}\ }\textbf {\bibinfo {volume} {19}},\ \bibinfo {pages} {4614}
  (\bibinfo {year} {2019})}\BibitemShut {NoStop}%
\bibitem [{\citenamefont {Huang}\ \emph
  {et~al.}(2020{\natexlab{a}})\citenamefont {Huang}, \citenamefont {Drost},
  \citenamefont {Senkpiel}, \citenamefont {Padurariu}, \citenamefont {Kubala},
  \citenamefont {Yeyati}, \citenamefont {Cuevas}, \citenamefont {Ankerhold},
  \citenamefont {Kern},\ and\ \citenamefont {Ast}}]{huang_quantum_2020}%
  \BibitemOpen
  \bibfield  {author} {\bibinfo {author} {\bibfnamefont {H.}~\bibnamefont
  {Huang}}, \bibinfo {author} {\bibfnamefont {R.}~\bibnamefont {Drost}},
  \bibinfo {author} {\bibfnamefont {J.}~\bibnamefont {Senkpiel}}, \bibinfo
  {author} {\bibfnamefont {C.}~\bibnamefont {Padurariu}}, \bibinfo {author}
  {\bibfnamefont {B.}~\bibnamefont {Kubala}}, \bibinfo {author} {\bibfnamefont
  {A.~L.}\ \bibnamefont {Yeyati}}, \bibinfo {author} {\bibfnamefont {J.~C.}\
  \bibnamefont {Cuevas}}, \bibinfo {author} {\bibfnamefont {J.}~\bibnamefont
  {Ankerhold}}, \bibinfo {author} {\bibfnamefont {K.}~\bibnamefont {Kern}}, \
  and\ \bibinfo {author} {\bibfnamefont {C.~R.}\ \bibnamefont {Ast}},\
  }\bibfield  {title} {{\selectlanguage {english}\emph {\bibinfo {title}
  {Quantum phase transitions and the role of impurity-substrate hybridization
  in {Yu}-{Shiba}-{Rusinov} states},}\ }}\href {\doibase
  10.1038/s42005-020-00469-0} {\bibfield  {journal} {\bibinfo  {journal}
  {Communications Physics}\ }\textbf {\bibinfo {volume} {3}},\ \bibinfo {pages}
  {1} (\bibinfo {year} {2020}{\natexlab{a}})}\BibitemShut {NoStop}%
\bibitem [{\citenamefont {Karan}\ \emph {et~al.}(2022)\citenamefont {Karan},
  \citenamefont {Huang}, \citenamefont {Padurariu}, \citenamefont {Kubala},
  \citenamefont {Theiler}, \citenamefont {Black-Schaffer}, \citenamefont
  {Morrás}, \citenamefont {Yeyati}, \citenamefont {Cuevas}, \citenamefont
  {Ankerhold}, \citenamefont {Kern},\ and\ \citenamefont
  {Ast}}]{karan_superconducting_2022}%
  \BibitemOpen
  \bibfield  {author} {\bibinfo {author} {\bibfnamefont {S.}~\bibnamefont
  {Karan}}, \bibinfo {author} {\bibfnamefont {H.}~\bibnamefont {Huang}},
  \bibinfo {author} {\bibfnamefont {C.}~\bibnamefont {Padurariu}}, \bibinfo
  {author} {\bibfnamefont {B.}~\bibnamefont {Kubala}}, \bibinfo {author}
  {\bibfnamefont {A.}~\bibnamefont {Theiler}}, \bibinfo {author} {\bibfnamefont
  {A.~M.}\ \bibnamefont {Black-Schaffer}}, \bibinfo {author} {\bibfnamefont
  {G.}~\bibnamefont {Morrás}}, \bibinfo {author} {\bibfnamefont {A.~L.}\
  \bibnamefont {Yeyati}}, \bibinfo {author} {\bibfnamefont {J.~C.}\
  \bibnamefont {Cuevas}}, \bibinfo {author} {\bibfnamefont {J.}~\bibnamefont
  {Ankerhold}}, \bibinfo {author} {\bibfnamefont {K.}~\bibnamefont {Kern}}, \
  and\ \bibinfo {author} {\bibfnamefont {C.~R.}\ \bibnamefont {Ast}},\
  }\bibfield  {title} {{\selectlanguage {english}\emph {\bibinfo {title}
  {Superconducting quantum interference at the atomic scale},}\ }}\href
  {\doibase 10.1038/s41567-022-01644-6} {\bibfield  {journal} {\bibinfo
  {journal} {Nature Physics}\ }\textbf {\bibinfo {volume} {18}},\ \bibinfo
  {pages} {893} (\bibinfo {year} {2022})}\BibitemShut {NoStop}%
\bibitem [{\citenamefont {Hatter}\ \emph {et~al.}(2015)\citenamefont {Hatter},
  \citenamefont {Heinrich}, \citenamefont {Ruby}, \citenamefont {Pascual},\
  and\ \citenamefont {Franke}}]{hatter_magnetic_2015}%
  \BibitemOpen
  \bibfield  {author} {\bibinfo {author} {\bibfnamefont {N.}~\bibnamefont
  {Hatter}}, \bibinfo {author} {\bibfnamefont {B.~W.}\ \bibnamefont
  {Heinrich}}, \bibinfo {author} {\bibfnamefont {M.}~\bibnamefont {Ruby}},
  \bibinfo {author} {\bibfnamefont {J.~I.}\ \bibnamefont {Pascual}}, \ and\
  \bibinfo {author} {\bibfnamefont {K.~J.}\ \bibnamefont {Franke}},\ }\bibfield
   {title} {{\selectlanguage {english}\emph {\bibinfo {title} {Magnetic
  anisotropy in {Shiba} bound states across a quantum phase transition},}\
  }}\href {\doibase 10.1038/ncomms9988} {\bibfield  {journal} {\bibinfo
  {journal} {Nature Communications}\ }\textbf {\bibinfo {volume} {6}},\
  \bibinfo {pages} {8988} (\bibinfo {year} {2015})}\BibitemShut {NoStop}%
\bibitem [{\citenamefont {Žitko}\ \emph {et~al.}(2011)\citenamefont {Žitko},
  \citenamefont {Bodensiek},\ and\ \citenamefont
  {Pruschke}}]{zitko_effects_2011}%
  \BibitemOpen
  \bibfield  {author} {\bibinfo {author} {\bibfnamefont {R.}~\bibnamefont
  {Žitko}}, \bibinfo {author} {\bibfnamefont {O.}~\bibnamefont {Bodensiek}}, \
  and\ \bibinfo {author} {\bibfnamefont {T.}~\bibnamefont {Pruschke}},\
  }\bibfield  {title} {\emph {\bibinfo {title} {Effects of magnetic anisotropy
  on the subgap excitations induced by quantum impurities in a superconducting
  host},}\ }\href {\doibase 10.1103/PhysRevB.83.054512} {\bibfield  {journal}
  {\bibinfo  {journal} {Physical Review B}\ }\textbf {\bibinfo {volume} {83}},\
  \bibinfo {pages} {054512} (\bibinfo {year} {2011})}\BibitemShut {NoStop}%
\bibitem [{\citenamefont {Cornils}\ \emph {et~al.}(2017)\citenamefont
  {Cornils}, \citenamefont {Kamlapure}, \citenamefont {Zhou}, \citenamefont
  {Pradhan}, \citenamefont {Khajetoorians}, \citenamefont {Fransson},
  \citenamefont {Wiebe},\ and\ \citenamefont
  {Wiesendanger}}]{cornils_spin-resolved_2017}%
  \BibitemOpen
  \bibfield  {author} {\bibinfo {author} {\bibfnamefont {L.}~\bibnamefont
  {Cornils}}, \bibinfo {author} {\bibfnamefont {A.}~\bibnamefont {Kamlapure}},
  \bibinfo {author} {\bibfnamefont {L.}~\bibnamefont {Zhou}}, \bibinfo {author}
  {\bibfnamefont {S.}~\bibnamefont {Pradhan}}, \bibinfo {author} {\bibfnamefont
  {A.}~\bibnamefont {Khajetoorians}}, \bibinfo {author} {\bibfnamefont
  {J.}~\bibnamefont {Fransson}}, \bibinfo {author} {\bibfnamefont
  {J.}~\bibnamefont {Wiebe}}, \ and\ \bibinfo {author} {\bibfnamefont
  {R.}~\bibnamefont {Wiesendanger}},\ }\bibfield  {title} {\emph {\bibinfo
  {title} {Spin-{Resolved} {Spectroscopy} of the {Yu}-{Shiba}-{Rusinov}
  {States} of {Individual} {Atoms}},}\ }\href {\doibase
  10.1103/PhysRevLett.119.197002} {\bibfield  {journal} {\bibinfo  {journal}
  {Physical Review Letters}\ }\textbf {\bibinfo {volume} {119}},\ \bibinfo
  {pages} {197002} (\bibinfo {year} {2017})}\BibitemShut {NoStop}%
\bibitem [{\citenamefont {Schneider}\ \emph {et~al.}(2021)\citenamefont
  {Schneider}, \citenamefont {Beck}, \citenamefont {Wiebe},\ and\ \citenamefont
  {Wiesendanger}}]{schneider_atomic-scale_2021}%
  \BibitemOpen
  \bibfield  {author} {\bibinfo {author} {\bibfnamefont {L.}~\bibnamefont
  {Schneider}}, \bibinfo {author} {\bibfnamefont {P.}~\bibnamefont {Beck}},
  \bibinfo {author} {\bibfnamefont {J.}~\bibnamefont {Wiebe}}, \ and\ \bibinfo
  {author} {\bibfnamefont {R.}~\bibnamefont {Wiesendanger}},\ }\bibfield
  {title} {{\selectlanguage {english}\emph {\bibinfo {title} {Atomic-scale
  spin-polarization maps using functionalized superconducting probes},}\
  }}\href {\doibase 10.1126/sciadv.abd7302} {\bibfield  {journal} {\bibinfo
  {journal} {Science Advances}\ }\textbf {\bibinfo {volume} {7}},\ \bibinfo
  {pages} {eabd7302} (\bibinfo {year} {2021})}\BibitemShut {NoStop}%
\bibitem [{\citenamefont {Machida}\ \emph {et~al.}(2022)\citenamefont
  {Machida}, \citenamefont {Nagai},\ and\ \citenamefont
  {Hanaguri}}]{machida_zeeman_2022}%
  \BibitemOpen
  \bibfield  {author} {\bibinfo {author} {\bibfnamefont {T.}~\bibnamefont
  {Machida}}, \bibinfo {author} {\bibfnamefont {Y.}~\bibnamefont {Nagai}}, \
  and\ \bibinfo {author} {\bibfnamefont {T.}~\bibnamefont {Hanaguri}},\
  }\bibfield  {title} {\emph {\bibinfo {title} {Zeeman effects on
  {Yu}-{Shiba}-{Rusinov} states},}\ }\href {\doibase
  10.1103/PhysRevResearch.4.033182} {\bibfield  {journal} {\bibinfo  {journal}
  {Physical Review Research}\ }\textbf {\bibinfo {volume} {4}},\ \bibinfo
  {pages} {033182} (\bibinfo {year} {2022})}\BibitemShut {NoStop}%
\bibitem [{\citenamefont {Lee}\ \emph {et~al.}(2014)\citenamefont {Lee},
  \citenamefont {Jiang}, \citenamefont {Houzet}, \citenamefont {Aguado},
  \citenamefont {Lieber},\ and\ \citenamefont
  {De~Franceschi}}]{lee_spin-resolved_2014}%
  \BibitemOpen
  \bibfield  {author} {\bibinfo {author} {\bibfnamefont {E.~J.~H.}\
  \bibnamefont {Lee}}, \bibinfo {author} {\bibfnamefont {X.}~\bibnamefont
  {Jiang}}, \bibinfo {author} {\bibfnamefont {M.}~\bibnamefont {Houzet}},
  \bibinfo {author} {\bibfnamefont {R.}~\bibnamefont {Aguado}}, \bibinfo
  {author} {\bibfnamefont {C.~M.}\ \bibnamefont {Lieber}}, \ and\ \bibinfo
  {author} {\bibfnamefont {S.}~\bibnamefont {De~Franceschi}},\ }\bibfield
  {title} {{\selectlanguage {english}\emph {\bibinfo {title} {Spin-resolved
  {Andreev} levels and parity crossings in hybrid
  superconductor–semiconductor nanostructures},}\ }}\href {\doibase
  10.1038/nnano.2013.267} {\bibfield  {journal} {\bibinfo  {journal} {Nature
  Nanotechnology}\ }\textbf {\bibinfo {volume} {9}},\ \bibinfo {pages} {79}
  (\bibinfo {year} {2014})}\BibitemShut {NoStop}%
\bibitem [{\citenamefont {Huang}\ \emph
  {et~al.}(2020{\natexlab{b}})\citenamefont {Huang}, \citenamefont {Padurariu},
  \citenamefont {Senkpiel}, \citenamefont {Drost}, \citenamefont {Yeyati},
  \citenamefont {Cuevas}, \citenamefont {Kubala}, \citenamefont {Ankerhold},
  \citenamefont {Kern},\ and\ \citenamefont {Ast}}]{huang_tunnelling_2020}%
  \BibitemOpen
  \bibfield  {author} {\bibinfo {author} {\bibfnamefont {H.}~\bibnamefont
  {Huang}}, \bibinfo {author} {\bibfnamefont {C.}~\bibnamefont {Padurariu}},
  \bibinfo {author} {\bibfnamefont {J.}~\bibnamefont {Senkpiel}}, \bibinfo
  {author} {\bibfnamefont {R.}~\bibnamefont {Drost}}, \bibinfo {author}
  {\bibfnamefont {A.~L.}\ \bibnamefont {Yeyati}}, \bibinfo {author}
  {\bibfnamefont {J.~C.}\ \bibnamefont {Cuevas}}, \bibinfo {author}
  {\bibfnamefont {B.}~\bibnamefont {Kubala}}, \bibinfo {author} {\bibfnamefont
  {J.}~\bibnamefont {Ankerhold}}, \bibinfo {author} {\bibfnamefont
  {K.}~\bibnamefont {Kern}}, \ and\ \bibinfo {author} {\bibfnamefont {C.~R.}\
  \bibnamefont {Ast}},\ }\bibfield  {title} {{\selectlanguage {english}\emph
  {\bibinfo {title} {Tunnelling dynamics between superconducting bound states
  at the atomic limit},}\ }}\href {\doibase 10.1038/s41567-020-0971-0}
  {\bibfield  {journal} {\bibinfo  {journal} {Nature Physics}\ }\textbf
  {\bibinfo {volume} {16}},\ \bibinfo {pages} {1227} (\bibinfo {year}
  {2020}{\natexlab{b}})}\BibitemShut {NoStop}%
\bibitem [{\citenamefont {Meservey}\ \emph {et~al.}(1970)\citenamefont
  {Meservey}, \citenamefont {Tedrow},\ and\ \citenamefont
  {Fulde}}]{meservey_magnetic_1970}%
  \BibitemOpen
  \bibfield  {author} {\bibinfo {author} {\bibfnamefont {R.}~\bibnamefont
  {Meservey}}, \bibinfo {author} {\bibfnamefont {P.~M.}\ \bibnamefont
  {Tedrow}}, \ and\ \bibinfo {author} {\bibfnamefont {P.}~\bibnamefont
  {Fulde}},\ }\bibfield  {title} {\emph {\bibinfo {title} {Magnetic field
  splitting of the quasiparticle states in superconducting aluminum films},}\
  }\href {\doibase 10.1103/PhysRevLett.25.1270} {\bibfield  {journal} {\bibinfo
   {journal} {Physical Review Letters}\ }\textbf {\bibinfo {volume} {25}},\
  \bibinfo {pages} {1270} (\bibinfo {year} {1970})}\BibitemShut {NoStop}%
\bibitem [{\citenamefont {Chen}\ \emph {et~al.}(2008)\citenamefont {Chen},
  \citenamefont {Doria},\ and\ \citenamefont {Peeters}}]{chen_vortices_2008}%
  \BibitemOpen
  \bibfield  {author} {\bibinfo {author} {\bibfnamefont {Y.}~\bibnamefont
  {Chen}}, \bibinfo {author} {\bibfnamefont {M.~M.}\ \bibnamefont {Doria}}, \
  and\ \bibinfo {author} {\bibfnamefont {F.~M.}\ \bibnamefont {Peeters}},\
  }\bibfield  {title} {\emph {\bibinfo {title} {Vortices in a mesoscopic cone:
  {A} superconducting tip in the presence of an applied field},}\ }\href
  {\doibase 10.1103/PhysRevB.77.054511} {\bibfield  {journal} {\bibinfo
  {journal} {Physical Review B}\ }\textbf {\bibinfo {volume} {77}},\ \bibinfo
  {pages} {054511} (\bibinfo {year} {2008})}\BibitemShut {NoStop}%
\bibitem [{\citenamefont {Eltschka}\ \emph {et~al.}(2014)\citenamefont
  {Eltschka}, \citenamefont {Jäck}, \citenamefont {Assig}, \citenamefont
  {Kondrashov}, \citenamefont {Skvortsov}, \citenamefont {Etzkorn},
  \citenamefont {Ast},\ and\ \citenamefont {Kern}}]{eltschka_probing_2014}%
  \BibitemOpen
  \bibfield  {author} {\bibinfo {author} {\bibfnamefont {M.}~\bibnamefont
  {Eltschka}}, \bibinfo {author} {\bibfnamefont {B.}~\bibnamefont {Jäck}},
  \bibinfo {author} {\bibfnamefont {M.}~\bibnamefont {Assig}}, \bibinfo
  {author} {\bibfnamefont {O.~V.}\ \bibnamefont {Kondrashov}}, \bibinfo
  {author} {\bibfnamefont {M.~A.}\ \bibnamefont {Skvortsov}}, \bibinfo {author}
  {\bibfnamefont {M.}~\bibnamefont {Etzkorn}}, \bibinfo {author} {\bibfnamefont
  {C.~R.}\ \bibnamefont {Ast}}, \ and\ \bibinfo {author} {\bibfnamefont
  {K.}~\bibnamefont {Kern}},\ }\bibfield  {title} {\emph {\bibinfo {title}
  {Probing absolute spin polarization at the nanoscale},}\ }\href {\doibase
  10.1021/nl5037947} {\bibfield  {journal} {\bibinfo  {journal} {Nano Lett.}\
  }\textbf {\bibinfo {volume} {14}},\ \bibinfo {pages} {7171} (\bibinfo {year}
  {2014})}\BibitemShut {NoStop}%
\bibitem [{\citenamefont {Ternes}\ \emph
  {et~al.}(2008{\natexlab{a}})\citenamefont {Ternes}, \citenamefont {Lutz},
  \citenamefont {Hirjibehedin}, \citenamefont {Giessibl},\ and\ \citenamefont
  {Heinrich}}]{ternes_force_2008}%
  \BibitemOpen
  \bibfield  {author} {\bibinfo {author} {\bibfnamefont {M.}~\bibnamefont
  {Ternes}}, \bibinfo {author} {\bibfnamefont {C.~P.}\ \bibnamefont {Lutz}},
  \bibinfo {author} {\bibfnamefont {C.~F.}\ \bibnamefont {Hirjibehedin}},
  \bibinfo {author} {\bibfnamefont {F.~J.}\ \bibnamefont {Giessibl}}, \ and\
  \bibinfo {author} {\bibfnamefont {A.~J.}\ \bibnamefont {Heinrich}},\
  }\bibfield  {title} {\emph {\bibinfo {title} {The force needed to move an
  atom on a surface},}\ }\href {\doibase 10.1126/science.1150288} {\bibfield
  {journal} {\bibinfo  {journal} {Science}\ }\textbf {\bibinfo {volume}
  {319}},\ \bibinfo {pages} {1066 } (\bibinfo {year}
  {2008}{\natexlab{a}})}\BibitemShut {NoStop}%
\bibitem [{\citenamefont {Ternes}\ \emph {et~al.}(2011)\citenamefont {Ternes},
  \citenamefont {Gonz\'{a}lez}, \citenamefont {Lutz}, \citenamefont {Hapala},
  \citenamefont {Giessibl}, \citenamefont {Jel\'{i}nek},\ and\ \citenamefont
  {Heinrich}}]{ternes_interplay_2011}%
  \BibitemOpen
  \bibfield  {author} {\bibinfo {author} {\bibfnamefont {M.}~\bibnamefont
  {Ternes}}, \bibinfo {author} {\bibfnamefont {C.}~\bibnamefont
  {Gonz\'{a}lez}}, \bibinfo {author} {\bibfnamefont {C.~P.}\ \bibnamefont
  {Lutz}}, \bibinfo {author} {\bibfnamefont {P.}~\bibnamefont {Hapala}},
  \bibinfo {author} {\bibfnamefont {F.~J.}\ \bibnamefont {Giessibl}}, \bibinfo
  {author} {\bibfnamefont {P.}~\bibnamefont {Jel\'{i}nek}}, \ and\ \bibinfo
  {author} {\bibfnamefont {A.~J.}\ \bibnamefont {Heinrich}},\ }\bibfield
  {title} {\emph {\bibinfo {title} {Interplay of conductance, force, and
  structural change in metallic point contacts},}\ }\href {\doibase
  10.1103/PhysRevLett.106.016802} {\bibfield  {journal} {\bibinfo  {journal}
  {Physical Review Letters}\ }\textbf {\bibinfo {volume} {106}},\ \bibinfo
  {pages} {016802} (\bibinfo {year} {2011})}\BibitemShut {NoStop}%
\bibitem [{\citenamefont {Sakurai}(1970)}]{sakurai_comments_1970}%
  \BibitemOpen
  \bibfield  {author} {\bibinfo {author} {\bibfnamefont {A.}~\bibnamefont
  {Sakurai}},\ }\bibfield  {title} {\emph {\bibinfo {title} {Comments on
  {Superconductors} with {Magnetic} {Impurities}},}\ }\href {\doibase
  10.1143/PTP.44.1472} {\bibfield  {journal} {\bibinfo  {journal} {Progress of
  Theoretical Physics}\ }\textbf {\bibinfo {volume} {44}},\ \bibinfo {pages}
  {1472} (\bibinfo {year} {1970})}\BibitemShut {NoStop}%
\bibitem [{sin()}]{sinf}%
  \BibitemOpen
  \href@noop {} {}\bibinfo {note} {See Supplementary Information.}\BibitemShut
  {Stop}%
\bibitem [{\citenamefont {Žitko}\ and\ \citenamefont
  {Pruschke}(2009)}]{zitko_energy_2009}%
  \BibitemOpen
  \bibfield  {author} {\bibinfo {author} {\bibfnamefont {R.}~\bibnamefont
  {Žitko}}\ and\ \bibinfo {author} {\bibfnamefont {T.}~\bibnamefont
  {Pruschke}},\ }\bibfield  {title} {\emph {\bibinfo {title} {Energy resolution
  and discretization artifacts in the numerical renormalization group},}\
  }\href {\doibase 10.1103/PhysRevB.79.085106} {\bibfield  {journal} {\bibinfo
  {journal} {Physical Review B}\ }\textbf {\bibinfo {volume} {79}},\ \bibinfo
  {pages} {085106} (\bibinfo {year} {2009})}\BibitemShut {NoStop}%
\bibitem [{\citenamefont {Kondo}(1964)}]{kondo_resistance_1964}%
  \BibitemOpen
  \bibfield  {author} {\bibinfo {author} {\bibfnamefont {J.}~\bibnamefont
  {Kondo}},\ }\bibfield  {title} {\emph {\bibinfo {title} {Resistance {Minimum}
  in {Dilute} {Magnetic} {Alloys}},}\ }\href {\doibase 10.1143/PTP.32.37}
  {\bibfield  {journal} {\bibinfo  {journal} {Progress of Theoretical Physics}\
  }\textbf {\bibinfo {volume} {32}},\ \bibinfo {pages} {37} (\bibinfo {year}
  {1964})}\BibitemShut {NoStop}%
\bibitem [{\citenamefont {Li}\ \emph {et~al.}(1998)\citenamefont {Li},
  \citenamefont {Schneider}, \citenamefont {Berndt},\ and\ \citenamefont
  {Delley}}]{li_kondo_1998}%
  \BibitemOpen
  \bibfield  {author} {\bibinfo {author} {\bibfnamefont {J.}~\bibnamefont
  {Li}}, \bibinfo {author} {\bibfnamefont {W.-D.}\ \bibnamefont {Schneider}},
  \bibinfo {author} {\bibfnamefont {R.}~\bibnamefont {Berndt}}, \ and\ \bibinfo
  {author} {\bibfnamefont {B.}~\bibnamefont {Delley}},\ }\bibfield  {title}
  {\emph {\bibinfo {title} {Kondo {Scattering} {Observed} at a {Single}
  {Magnetic} {Impurity}},}\ }\href {\doibase 10.1103/PhysRevLett.80.2893}
  {\bibfield  {journal} {\bibinfo  {journal} {Physical Review Letters}\
  }\textbf {\bibinfo {volume} {80}},\ \bibinfo {pages} {2893} (\bibinfo {year}
  {1998})}\BibitemShut {NoStop}%
\bibitem [{\citenamefont {Madhavan}\ \emph {et~al.}(1998)\citenamefont
  {Madhavan}, \citenamefont {Chen}, \citenamefont {Jamneala}, \citenamefont
  {Crommie},\ and\ \citenamefont {Wingreen}}]{madhavan_tunneling_1998}%
  \BibitemOpen
  \bibfield  {author} {\bibinfo {author} {\bibfnamefont {V.}~\bibnamefont
  {Madhavan}}, \bibinfo {author} {\bibfnamefont {W.}~\bibnamefont {Chen}},
  \bibinfo {author} {\bibfnamefont {T.}~\bibnamefont {Jamneala}}, \bibinfo
  {author} {\bibfnamefont {M.~F.}\ \bibnamefont {Crommie}}, \ and\ \bibinfo
  {author} {\bibfnamefont {N.~S.}\ \bibnamefont {Wingreen}},\ }\bibfield
  {title} {\emph {\bibinfo {title} {Tunneling into a {Single} {Magnetic}
  {Atom}: {Spectroscopic} {Evidence} of the {Kondo} {Resonance}},}\ }\href
  {\doibase 10.1126/science.280.5363.567} {\bibfield  {journal} {\bibinfo
  {journal} {Science}\ }\textbf {\bibinfo {volume} {280}},\ \bibinfo {pages}
  {567} (\bibinfo {year} {1998})}\BibitemShut {NoStop}%
\bibitem [{\citenamefont {Ternes}\ \emph
  {et~al.}(2008{\natexlab{b}})\citenamefont {Ternes}, \citenamefont
  {Heinrich},\ and\ \citenamefont {Schneider}}]{ternes_spectroscopic_2008}%
  \BibitemOpen
  \bibfield  {author} {\bibinfo {author} {\bibfnamefont {M.}~\bibnamefont
  {Ternes}}, \bibinfo {author} {\bibfnamefont {A.~J.}\ \bibnamefont
  {Heinrich}}, \ and\ \bibinfo {author} {\bibfnamefont {W.-D.}\ \bibnamefont
  {Schneider}},\ }\bibfield  {title} {{\selectlanguage {english}\emph {\bibinfo
  {title} {Spectroscopic manifestations of the {Kondo} effect on single
  adatoms},}\ }}\href {\doibase 10.1088/0953-8984/21/5/053001} {\bibfield
  {journal} {\bibinfo  {journal} {Journal of Physics: Condensed Matter}\
  }\textbf {\bibinfo {volume} {21}},\ \bibinfo {pages} {053001} (\bibinfo
  {year} {2008}{\natexlab{b}})}\BibitemShut {NoStop}%
\bibitem [{\citenamefont {Huang}\ \emph {et~al.}(2022)\citenamefont {Huang},
  \citenamefont {Karan}, \citenamefont {Padurariu}, \citenamefont {Kubala},
  \citenamefont {Cuevas}, \citenamefont {Ankerhold}, \citenamefont {Kern},\
  and\ \citenamefont {Ast}}]{huang_universal_2022}%
  \BibitemOpen
  \bibfield  {author} {\bibinfo {author} {\bibfnamefont {H.}~\bibnamefont
  {Huang}}, \bibinfo {author} {\bibfnamefont {S.}~\bibnamefont {Karan}},
  \bibinfo {author} {\bibfnamefont {C.}~\bibnamefont {Padurariu}}, \bibinfo
  {author} {\bibfnamefont {B.}~\bibnamefont {Kubala}}, \bibinfo {author}
  {\bibfnamefont {J.~C.}\ \bibnamefont {Cuevas}}, \bibinfo {author}
  {\bibfnamefont {J.}~\bibnamefont {Ankerhold}}, \bibinfo {author}
  {\bibfnamefont {K.}~\bibnamefont {Kern}}, \ and\ \bibinfo {author}
  {\bibfnamefont {C.~R.}\ \bibnamefont {Ast}},\ }\bibfield  {title} {\emph
  {\bibinfo {title} {Universal scaling of tunable {Yu-Shiba-Rusinov} states
  across the quantum phase transition},}\ }\href {\doibase
  10.48550/arXiv.2212.11332} {\bibfield  {journal} {\bibinfo  {journal}
  {arXiv:2212.11332}\ } (\bibinfo {year} {2022})}\BibitemShut {NoStop}%
\bibitem [{\citenamefont {Cuevas}\ \emph {et~al.}(1998)\citenamefont {Cuevas},
  \citenamefont {Levy~Yeyati}, \citenamefont {Mart{\'i}n-Rodero}, \citenamefont
  {Rubio~Bollinger}, \citenamefont {Untiedt},\ and\ \citenamefont
  {Agra{\"\i}t}}]{cuevas_evolution_1998}%
  \BibitemOpen
  \bibfield  {author} {\bibinfo {author} {\bibfnamefont {J.~C.}\ \bibnamefont
  {Cuevas}}, \bibinfo {author} {\bibfnamefont {A.}~\bibnamefont {Levy~Yeyati}},
  \bibinfo {author} {\bibfnamefont {A.}~\bibnamefont {Mart{\'i}n-Rodero}},
  \bibinfo {author} {\bibfnamefont {G.}~\bibnamefont {Rubio~Bollinger}},
  \bibinfo {author} {\bibfnamefont {C.}~\bibnamefont {Untiedt}}, \ and\
  \bibinfo {author} {\bibfnamefont {N.}~\bibnamefont {Agra{\"\i}t}},\
  }\bibfield  {title} {\emph {\bibinfo {title} {Evolution of {Conducting}
  {Channels} in {Metallic} {Atomic} {Contacts} under {Elastic}
  {Deformation}},}\ }\href {\doibase 10.1103/PhysRevLett.81.2990} {\bibfield
  {journal} {\bibinfo  {journal} {Physical Review Letters}\ }\textbf {\bibinfo
  {volume} {81}},\ \bibinfo {pages} {2990} (\bibinfo {year}
  {1998})}\BibitemShut {NoStop}%
\bibitem [{\citenamefont {Hewson}\ \emph {et~al.}(2005)\citenamefont {Hewson},
  \citenamefont {Bauer},\ and\ \citenamefont
  {Oguri}}]{hewson_non-equilibrium_2005}%
  \BibitemOpen
  \bibfield  {author} {\bibinfo {author} {\bibfnamefont {A.~C.}\ \bibnamefont
  {Hewson}}, \bibinfo {author} {\bibfnamefont {J.}~\bibnamefont {Bauer}}, \
  and\ \bibinfo {author} {\bibfnamefont {A.}~\bibnamefont {Oguri}},\ }\bibfield
   {title} {{\selectlanguage {english}\emph {\bibinfo {title} {Non-equilibrium
  differential conductance through a quantum dot in a magnetic field},}\
  }}\href {\doibase 10.1088/0953-8984/17/35/008} {\bibfield  {journal}
  {\bibinfo  {journal} {Journal of Physics: Condensed Matter}\ }\textbf
  {\bibinfo {volume} {17}},\ \bibinfo {pages} {5413} (\bibinfo {year}
  {2005})}\BibitemShut {NoStop}%
\bibitem [{\citenamefont {{\v{Z}}itko}\ \emph {et~al.}(2009)\citenamefont
  {{\v{Z}}itko}, \citenamefont {Peters},\ and\ \citenamefont
  {Pruschke}}]{zitko2009splitting}%
  \BibitemOpen
  \bibfield  {author} {\bibinfo {author} {\bibfnamefont {R.}~\bibnamefont
  {{\v{Z}}itko}}, \bibinfo {author} {\bibfnamefont {R.}~\bibnamefont {Peters}},
  \ and\ \bibinfo {author} {\bibfnamefont {T.}~\bibnamefont {Pruschke}},\
  }\bibfield  {title} {\emph {\bibinfo {title} {Splitting of the {Kondo}
  resonance in anisotropic magnetic impurities on surfaces},}\ }\href {\doibase
  10.1103/PhysRevB.78.224404} {\bibfield  {journal} {\bibinfo  {journal} {New
  J. Phys.}\ }\textbf {\bibinfo {volume} {11}},\ \bibinfo {pages} {053003}
  (\bibinfo {year} {2009})}\BibitemShut {NoStop}%
\bibitem [{\citenamefont {Kretinin}\ \emph {et~al.}(2011)\citenamefont
  {Kretinin}, \citenamefont {Shtrikman}, \citenamefont {Goldhaber-Gordon},
  \citenamefont {Hanl}, \citenamefont {Weichselbaum}, \citenamefont {von
  Delft}, \citenamefont {Costi},\ and\ \citenamefont
  {Mahalu}}]{kretinin_spin-frac12_2011}%
  \BibitemOpen
  \bibfield  {author} {\bibinfo {author} {\bibfnamefont {A.~V.}\ \bibnamefont
  {Kretinin}}, \bibinfo {author} {\bibfnamefont {H.}~\bibnamefont {Shtrikman}},
  \bibinfo {author} {\bibfnamefont {D.}~\bibnamefont {Goldhaber-Gordon}},
  \bibinfo {author} {\bibfnamefont {M.}~\bibnamefont {Hanl}}, \bibinfo {author}
  {\bibfnamefont {A.}~\bibnamefont {Weichselbaum}}, \bibinfo {author}
  {\bibfnamefont {J.}~\bibnamefont {von Delft}}, \bibinfo {author}
  {\bibfnamefont {T.}~\bibnamefont {Costi}}, \ and\ \bibinfo {author}
  {\bibfnamefont {D.}~\bibnamefont {Mahalu}},\ }\bibfield  {title} {\emph
  {\bibinfo {title} {Spin-1/2 {Kondo} effect in an {InAs} nanowire quantum dot:
  {Unitary} limit, conductance scaling, and {Zeeman} splitting},}\ }\href
  {\doibase 10.1103/PhysRevB.84.245316} {\bibfield  {journal} {\bibinfo
  {journal} {Physical Review B}\ }\textbf {\bibinfo {volume} {84}},\ \bibinfo
  {pages} {245316} (\bibinfo {year} {2011})}\BibitemShut {NoStop}%
\bibitem [{\citenamefont {Satori}\ \emph {et~al.}(1992)\citenamefont {Satori},
  \citenamefont {Shiba}, \citenamefont {Sakai},\ and\ \citenamefont
  {Shimizu}}]{satori_numerical_1992}%
  \BibitemOpen
  \bibfield  {author} {\bibinfo {author} {\bibfnamefont {K.}~\bibnamefont
  {Satori}}, \bibinfo {author} {\bibfnamefont {H.}~\bibnamefont {Shiba}},
  \bibinfo {author} {\bibfnamefont {O.}~\bibnamefont {Sakai}}, \ and\ \bibinfo
  {author} {\bibfnamefont {Y.}~\bibnamefont {Shimizu}},\ }\bibfield  {title}
  {\emph {\bibinfo {title} {Numerical {Renormalization} {Group} {Study} of
  {Magnetic} {Impurities} in {Superconductors}},}\ }\href {\doibase
  10.1143/JPSJ.61.3239} {\bibfield  {journal} {\bibinfo  {journal} {Journal of
  the Physical Society of Japan}\ }\textbf {\bibinfo {volume} {61}},\ \bibinfo
  {pages} {3239} (\bibinfo {year} {1992})}\BibitemShut {NoStop}%
\bibitem [{\citenamefont {Yoshioka}\ and\ \citenamefont
  {Ohashi}(2000)}]{yoshioka_numerical_2000}%
  \BibitemOpen
  \bibfield  {author} {\bibinfo {author} {\bibfnamefont {T.}~\bibnamefont
  {Yoshioka}}\ and\ \bibinfo {author} {\bibfnamefont {Y.}~\bibnamefont
  {Ohashi}},\ }\bibfield  {title} {\emph {\bibinfo {title} {Numerical
  {Renormalization} {Group} {Studies} on {Single} {Impurity} {Anderson} {Model}
  in {Superconductivity}: {A} {Unified} {Treatment} of {Magnetic},
  {Nonmagnetic} {Impurities}, and {Resonance} {Scattering}},}\ }\href {\doibase
  10.1143/JPSJ.69.1812} {\bibfield  {journal} {\bibinfo  {journal} {Journal of
  the Physical Society of Japan}\ }\textbf {\bibinfo {volume} {69}},\ \bibinfo
  {pages} {1812} (\bibinfo {year} {2000})}\BibitemShut {NoStop}%
\bibitem [{\citenamefont {Bulla}\ \emph {et~al.}(2008)\citenamefont {Bulla},
  \citenamefont {Costi},\ and\ \citenamefont
  {Pruschke}}]{bulla_numerical_2008}%
  \BibitemOpen
  \bibfield  {author} {\bibinfo {author} {\bibfnamefont {R.}~\bibnamefont
  {Bulla}}, \bibinfo {author} {\bibfnamefont {T.~A.}\ \bibnamefont {Costi}}, \
  and\ \bibinfo {author} {\bibfnamefont {T.}~\bibnamefont {Pruschke}},\
  }\bibfield  {title} {\emph {\bibinfo {title} {Numerical renormalization group
  method for quantum impurity systems},}\ }\href {\doibase
  10.1103/RevModPhys.80.395} {\bibfield  {journal} {\bibinfo  {journal}
  {Reviews of Modern Physics}\ }\textbf {\bibinfo {volume} {80}},\ \bibinfo
  {pages} {395} (\bibinfo {year} {2008})}\BibitemShut {NoStop}%
\bibitem [{\citenamefont {von Oppen}\ and\ \citenamefont
  {Franke}(2021)}]{oppen_yu-shiba-rusinov_2021}%
  \BibitemOpen
  \bibfield  {author} {\bibinfo {author} {\bibfnamefont {F.}~\bibnamefont {von
  Oppen}}\ and\ \bibinfo {author} {\bibfnamefont {K.~J.}\ \bibnamefont
  {Franke}},\ }\bibfield  {title} {\emph {\bibinfo {title}
  {Yu-{Shiba}-{Rusinov} states in real metals},}\ }\href {\doibase
  10.1103/PhysRevB.103.205424} {\bibfield  {journal} {\bibinfo  {journal}
  {Physical Review B}\ }\textbf {\bibinfo {volume} {103}},\ \bibinfo {pages}
  {205424} (\bibinfo {year} {2021})}\BibitemShut {NoStop}%
\bibitem [{\citenamefont {Koller}\ \emph {et~al.}(2001)\citenamefont {Koller},
  \citenamefont {Bergermayer}, \citenamefont {Kresse}, \citenamefont
  {Hebenstreit}, \citenamefont {Konvicka}, \citenamefont {Schmid},
  \citenamefont {Podloucky},\ and\ \citenamefont
  {Varga}}]{koller_structure_2001}%
  \BibitemOpen
  \bibfield  {author} {\bibinfo {author} {\bibfnamefont {R.}~\bibnamefont
  {Koller}}, \bibinfo {author} {\bibfnamefont {W.}~\bibnamefont {Bergermayer}},
  \bibinfo {author} {\bibfnamefont {G.}~\bibnamefont {Kresse}}, \bibinfo
  {author} {\bibfnamefont {E.~L.~D.}\ \bibnamefont {Hebenstreit}}, \bibinfo
  {author} {\bibfnamefont {C.}~\bibnamefont {Konvicka}}, \bibinfo {author}
  {\bibfnamefont {M.}~\bibnamefont {Schmid}}, \bibinfo {author} {\bibfnamefont
  {R.}~\bibnamefont {Podloucky}}, \ and\ \bibinfo {author} {\bibfnamefont
  {P.}~\bibnamefont {Varga}},\ }\bibfield  {title} {{\selectlanguage
  {english}\emph {\bibinfo {title} {The structure of the oxygen induced (1×5)
  reconstruction of {V}(100)},}\ }}\href {\doibase
  10.1016/S0039-6028(01)00978-5} {\bibfield  {journal} {\bibinfo  {journal}
  {Surface Science}\ }\textbf {\bibinfo {volume} {480}},\ \bibinfo {pages} {11}
  (\bibinfo {year} {2001})}\BibitemShut {NoStop}%
\bibitem [{\citenamefont {Kralj}\ \emph {et~al.}(2003)\citenamefont {Kralj},
  \citenamefont {Pervan}, \citenamefont {Milun}, \citenamefont {Wandelt},
  \citenamefont {Mandrino},\ and\ \citenamefont {Jenko}}]{kralj_hraes_2003}%
  \BibitemOpen
  \bibfield  {author} {\bibinfo {author} {\bibfnamefont {M.}~\bibnamefont
  {Kralj}}, \bibinfo {author} {\bibfnamefont {P.}~\bibnamefont {Pervan}},
  \bibinfo {author} {\bibfnamefont {M.}~\bibnamefont {Milun}}, \bibinfo
  {author} {\bibfnamefont {K.}~\bibnamefont {Wandelt}}, \bibinfo {author}
  {\bibfnamefont {D.}~\bibnamefont {Mandrino}}, \ and\ \bibinfo {author}
  {\bibfnamefont {M.}~\bibnamefont {Jenko}},\ }\bibfield  {title}
  {{\selectlanguage {english}\emph {\bibinfo {title} {{HRAES}, {STM} and
  {ARUPS} study of (5×1) reconstructed {V}(100)},}\ }}\href {\doibase
  10.1016/S0039-6028(02)02647-X} {\bibfield  {journal} {\bibinfo  {journal}
  {Surface Science}\ }\textbf {\bibinfo {volume} {526}},\ \bibinfo {pages}
  {166} (\bibinfo {year} {2003})}\BibitemShut {NoStop}%
\bibitem [{\citenamefont {Schrieffer}\ and\ \citenamefont
  {Wolff}(1966)}]{schrieffer_relation_1966}%
  \BibitemOpen
  \bibfield  {author} {\bibinfo {author} {\bibfnamefont {J.~R.}\ \bibnamefont
  {Schrieffer}}\ and\ \bibinfo {author} {\bibfnamefont {P.~A.}\ \bibnamefont
  {Wolff}},\ }\bibfield  {title} {\emph {\bibinfo {title} {Relation between the
  {Anderson} and {Kondo} {Hamiltonians}},}\ }\href {\doibase
  10.1103/PhysRev.149.491} {\bibfield  {journal} {\bibinfo  {journal} {Physical
  Review}\ }\textbf {\bibinfo {volume} {149}},\ \bibinfo {pages} {491}
  (\bibinfo {year} {1966})}\BibitemShut {NoStop}%
\bibitem [{\citenamefont {Kadlecov{\'a}}\ \emph {et~al.}(2019)\citenamefont
  {Kadlecov{\'a}}, \citenamefont {{\v{Z}}onda}, \citenamefont {Pokorn{\`y}},\
  and\ \citenamefont {Novotn{\`y}}}]{kadlecova2019practical}%
  \BibitemOpen
  \bibfield  {author} {\bibinfo {author} {\bibfnamefont {A.}~\bibnamefont
  {Kadlecov{\'a}}}, \bibinfo {author} {\bibfnamefont {M.}~\bibnamefont
  {{\v{Z}}onda}}, \bibinfo {author} {\bibfnamefont {V.}~\bibnamefont
  {Pokorn{\`y}}}, \ and\ \bibinfo {author} {\bibfnamefont {T.}~\bibnamefont
  {Novotn{\`y}}},\ }\bibfield  {title} {\emph {\bibinfo {title} {Practical
  guide to quantum phase transitions in quantum-dot-based tunable {Josephson}
  junctions},}\ }\href {\doibase 10.1103/PhysRevApplied.11.044094} {\bibfield
  {journal} {\bibinfo  {journal} {Physical Review Applied}\ }\textbf {\bibinfo
  {volume} {11}},\ \bibinfo {pages} {044094} (\bibinfo {year}
  {2019})}\BibitemShut {NoStop}%
\bibitem [{\citenamefont {{\v{Z}}itko}(2007)}]{zitko2007many}%
  \BibitemOpen
  \bibfield  {author} {\bibinfo {author} {\bibfnamefont {R.}~\bibnamefont
  {{\v{Z}}itko}},\ }\emph {\bibinfo {title} {Many-particle effects in resonant
  tunneling of electrons through nanostructures}},\ \href
  {https://www.dlib.si/details/URN:NBN:SI:DOC-NSOD6JSH} {Ph.D. thesis},\
  \bibinfo  {school} {University of Ljubljana}, \bibinfo {address} {Ljubljana}
  (\bibinfo {year} {2007})\BibitemShut {NoStop}%
\end{thebibliography}

\begin{thebibliography}{5}%
\makeatletter
\providecommand \@ifxundefined [1]{%
 \@ifx{#1\undefined}
}%
\providecommand \@ifnum [1]{%
 \ifnum #1\expandafter \@firstoftwo
 \else \expandafter \@secondoftwo
 \fi
}%
\providecommand \@ifx [1]{%
 \ifx #1\expandafter \@firstoftwo
 \else \expandafter \@secondoftwo
 \fi
}%
\providecommand \natexlab [1]{#1}%
\providecommand \emph  [1]{``#1''}%
\providecommand \bibnamefont  [1]{#1}%
\providecommand \bibfnamefont [1]{#1}%
\providecommand \citenamefont [1]{#1}%
\providecommand \href@noop [0]{\@secondoftwo}%
\providecommand \href [0]{\begingroup \@sanitize@url \@href}%
\providecommand \@href[1]{\@@startlink{#1}\@@href}%
\providecommand \@@href[1]{\endgroup#1\@@endlink}%
\providecommand \@sanitize@url [0]{\catcode `\\12\catcode `\$12\catcode
  `\&12\catcode `\#12\catcode `\^12\catcode `\_12\catcode `\%12\relax}%
\providecommand \@@startlink[1]{}%
\providecommand \@@endlink[0]{}%
\providecommand \url  [0]{\begingroup\@sanitize@url \@url }%
\providecommand \@url [1]{\endgroup\@href {#1}{\urlprefix }}%
\providecommand \urlprefix  [0]{URL }%
\providecommand \Eprint [0]{\href }%
\providecommand \doibase [0]{http://dx.doi.org/}%
\providecommand \selectlanguage [0]{\@gobble}%
\providecommand \bibinfo  [0]{\@secondoftwo}%
\providecommand \bibfield  [0]{\@secondoftwo}%
\providecommand \translation [1]{[#1]}%
\providecommand \BibitemOpen [0]{}%
\providecommand \bibitemStop [0]{}%
\providecommand \bibitemNoStop [0]{.\EOS\space}%
\providecommand \EOS [0]{\spacefactor3000\relax}%
\providecommand \BibitemShut  [1]{\csname bibitem#1\endcsname}%
\let\auto@bib@innerbib\@empty
%</preamble>
\bibitem [{\citenamefont {Anderson}(1959)}]{si_Anderson1959}%
  \BibitemOpen
  \bibfield  {author} {\bibinfo {author} {\bibfnamefont {P.}~\bibnamefont
  {Anderson}},\ }\bibfield  {title} {\emph {\bibinfo {title} {Theory of dirty
  superconductors},}\ }\href {\doibase
  https://doi.org/10.1016/0022-3697(59)90036-8} {\bibfield  {journal} {\bibinfo
   {journal} {Journal of Physics and Chemistry of Solids}\ }\textbf {\bibinfo
  {volume} {11}},\ \bibinfo {pages} {26} (\bibinfo {year} {1959})}\BibitemShut
  {NoStop}%
\bibitem [{\citenamefont {\ifmmode~\check{z}\else \v{Z}\fi{}itko}\ \emph
  {et~al.}(2011)\citenamefont {\ifmmode~\check{z}\else \v{Z}\fi{}itko},
  \citenamefont {Bodensiek},\ and\ \citenamefont {Pruschke}}]{si_Zitko2011}%
  \BibitemOpen
  \bibfield  {author} {\bibinfo {author} {\bibfnamefont {R.}~\bibnamefont
  {\ifmmode~\check{z}\else \v{Z}\fi{}itko}}, \bibinfo {author} {\bibfnamefont
  {O.}~\bibnamefont {Bodensiek}}, \ and\ \bibinfo {author} {\bibfnamefont
  {T.}~\bibnamefont {Pruschke}},\ }\bibfield  {title} {\emph {\bibinfo {title}
  {Effects of magnetic anisotropy on the subgap excitations induced by quantum
  impurities in a superconducting host},}\ }\href {\doibase
  10.1103/PhysRevB.83.054512} {\bibfield  {journal} {\bibinfo  {journal} {Phys.
  Rev. B}\ }\textbf {\bibinfo {volume} {83}},\ \bibinfo {pages} {054512}
  (\bibinfo {year} {2011})}\BibitemShut {NoStop}%
\bibitem [{\citenamefont {Huang}\ \emph {et~al.}(2020)\citenamefont {Huang},
  \citenamefont {Drost}, \citenamefont {Senkpiel}, \citenamefont {Padurariu},
  \citenamefont {Kubala}, \citenamefont {Yeyati}, \citenamefont {Cuevas},
  \citenamefont {Ankerhold}, \citenamefont {Kern},\ and\ \citenamefont
  {Ast}}]{si_Huang2020}%
  \BibitemOpen
  \bibfield  {author} {\bibinfo {author} {\bibfnamefont {H.}~\bibnamefont
  {Huang}}, \bibinfo {author} {\bibfnamefont {R.}~\bibnamefont {Drost}},
  \bibinfo {author} {\bibfnamefont {J.}~\bibnamefont {Senkpiel}}, \bibinfo
  {author} {\bibfnamefont {C.}~\bibnamefont {Padurariu}}, \bibinfo {author}
  {\bibfnamefont {B.}~\bibnamefont {Kubala}}, \bibinfo {author} {\bibfnamefont
  {A.~L.}\ \bibnamefont {Yeyati}}, \bibinfo {author} {\bibfnamefont {J.~C.}\
  \bibnamefont {Cuevas}}, \bibinfo {author} {\bibfnamefont {J.}~\bibnamefont
  {Ankerhold}}, \bibinfo {author} {\bibfnamefont {K.}~\bibnamefont {Kern}}, \
  and\ \bibinfo {author} {\bibfnamefont {C.~R.}\ \bibnamefont {Ast}},\
  }\bibfield  {title} {\emph {\bibinfo {title} {Quantum phase transitions and
  the role of impurity-substrate hybridization in {Yu-Shiba-Rusinov} states},}\
  }\href {\doibase 10.1038/s42005-020-00469-0} {\bibfield  {journal} {\bibinfo
  {journal} {Communications Physics}\ }\textbf {\bibinfo {volume} {3}},\
  \bibinfo {pages} {199} (\bibinfo {year} {2020})}\BibitemShut {NoStop}%
\bibitem [{\citenamefont {Machida}\ \emph {et~al.}(2022)\citenamefont
  {Machida}, \citenamefont {Nagai},\ and\ \citenamefont
  {Hanaguri}}]{si_Machida2022}%
  \BibitemOpen
  \bibfield  {author} {\bibinfo {author} {\bibfnamefont {T.}~\bibnamefont
  {Machida}}, \bibinfo {author} {\bibfnamefont {Y.}~\bibnamefont {Nagai}}, \
  and\ \bibinfo {author} {\bibfnamefont {T.}~\bibnamefont {Hanaguri}},\
  }\bibfield  {title} {\emph {\bibinfo {title} {Zeeman effects on
  {Yu-Shiba-Rusinov} states},}\ }\href {\doibase
  10.1103/PhysRevResearch.4.033182} {\bibfield  {journal} {\bibinfo  {journal}
  {Phys. Rev. Res.}\ }\textbf {\bibinfo {volume} {4}},\ \bibinfo {pages}
  {033182} (\bibinfo {year} {2022})}\BibitemShut {NoStop}%
\bibitem [{\citenamefont {van Gerven~Oei}\ \emph {et~al.}(2017)\citenamefont
  {van Gerven~Oei}, \citenamefont {Tanaskovi\ifmmode~\acute{c}\else
  \'{c}\fi{}},\ and\ \citenamefont {\ifmmode~\check{Z}\else
  \v{Z}\fi{}itko}}]{si_van_Gerven_Oei_2017}%
  \BibitemOpen
  \bibfield  {author} {\bibinfo {author} {\bibfnamefont {W.-V.}\ \bibnamefont
  {van Gerven~Oei}}, \bibinfo {author} {\bibfnamefont {D.}~\bibnamefont
  {Tanaskovi\ifmmode~\acute{c}\else \'{c}\fi{}}}, \ and\ \bibinfo {author}
  {\bibfnamefont {R.}~\bibnamefont {\ifmmode~\check{Z}\else \v{Z}\fi{}itko}},\
  }\bibfield  {title} {\emph {\bibinfo {title} {Magnetic impurities in
  spin-split superconductors},}\ }\href {\doibase 10.1103/PhysRevB.95.085115}
  {\bibfield  {journal} {\bibinfo  {journal} {Phys. Rev. B}\ }\textbf {\bibinfo
  {volume} {95}},\ \bibinfo {pages} {085115} (\bibinfo {year}
  {2017})}\BibitemShut {NoStop}%
\end{thebibliography}
\end{document}